# CMB-S4: Iterative Internal Delensing and *r* Constraints

Sebastian Belkner[1], Julien Carron[1], Louis Legrand[1], Caterina Umiltà[2], Clem Pryke[3], and Colin Bischoff[4]
CMB-S4 Collaboration
[1] Université de Genève, Département de Physique Théorique et CAP, 24 Quai Ansermet, CH-1211 Genève 4, Switzerland; sebastian.belkner@unige.ch, julien.carron@unige.ch
[2] Department of Physics, University of Illinois Urbana-Champaign, 1110 West Green Street, Urbana, IL 61801, USA
[3] Department of Physics, University of Minnesota, Minneapolis, MN 55455, USA
[4] Department of Physics, University of Cincinnati, Cincinnati, OH 45221, USA

## Abstract

The tightest constraints on the tensor-to-scalar ratio *r* can only be obtained after removing a substantial fraction of the lensing *B*-mode sample variance. The planned Cosmic Microwave Background (CMB)-S4 experiment (cmb-s4.org) will remove the lensing *B*-mode signal internally by reconstructing the gravitational lenses from high-resolution observations. We document here a first lensing reconstruction pipeline able to achieve this optimally for arbitrary sky coverage. We make it part of a map-based framework to test CMB-S4 delensing performance and its constraining power on *r*, including inhomogeneous noise and two non-Gaussian Galactic polarized foreground models. The framework performs component separation of the high-resolution maps, followed by the construction of lensing *B*-mode templates, which are then included in a parametric small-aperture map cross-spectra-based likelihood for *r*. We find that the lensing reconstruction and framework achieve the expected performance, compatible with the target $\sigma(r) \simeq 5 \cdot 10^{-4}$ in the absence of a tensor signal, after an effective removal of 92%–93% of the lensing *B*-mode variance, depending on the simulation set. The code for the lensing reconstruction can also be used for cross-correlation studies with large-scale structures, lensing spectrum reconstruction, cluster lensing, or other CMB lensing-related purposes. As part of our tests, we also demonstrate the joint optimal reconstruction of the lensing potential with the lensing curl potential mode at second order in the density fluctuations.

*Unified Astronomy Thesaurus concepts:* Cosmic background radiation (317); Cosmological parameters (339); Gravitational waves (678); Cosmic inflation (319)

## 1. Introduction

According to the inflationary paradigm, the same mechanism responsible for the seeds of structure in the Universe also may produce a sizeable background of primordial gravitational waves (PGWs). These waves, or tensor perturbations, leave a signature on the polarization of the cosmic microwave background (CMB), which is potentially observable with current technology (Kamionkowski et al. 1997; Seljak 1997; Seljak & Zaldarriaga 1997). So far, several fundamental predictions of the simplest inflationary models have been confirmed by observations, including a nearly but not exactly scale-invariant spectrum of almost completely Gaussian and adiabatic scalar perturbations (Planck Collaboration et al. 2020a). Detection of the tensor signal would provide us a closer view than ever of the earliest moments of our Universe, and would provide strong support for large-field inflation, with important implications for our understanding of the inflationary phase and physics at very high energies (see Kamionkowski & Kovetz 2016; Achúcarro et al. 2022 or Lyth & Riotto 1999; Baumann & McAllister 2015, for reviews).

The tensor signal is most distinctive on the degree scales of the CMB polarization *B* mode, owing to the absence of first-order scalar contributions and their associated cosmic variance (Hu & White 1997; Kamionkowski et al. 1997; Seljak & Zaldarriaga 1997). Several experiments have been conducted or are currently undertaking measurements of the *B*-mode power. The primordial component is usually parameterized by the tensor-to-scalar ratio *r*. The tightest constraint to date[5] is $r_{0.05} < 0.032$ at 95% CL., and comes from a combination of Planck and BICEP/Keck measurements (Ade et al. 2021; Tristram et al. 2022).

The limiting factor in this constraint is now the leading nonlinear effect in the CMB: gravitational lensing. During their journey across the Universe, CMB photons are deflected by the pull of large-scale structures. The light is deflected by a few arcminutes at all frequencies by large-scale lenses distributed across a wide range of redshifts (Lewis & Challinor 2006). In particular, this distorts the primordial polarization pattern of the CMB slightly: a pure-*E* polarization pattern will produce a sizeable amount of *B* modes after lensing (Zaldarriaga & Seljak 1998; Kesden et al. 2002; Knox & Song 2002). On the scales relevant for *r* measurements, this lensing effect appears as a white-noise power with an amplitude close to 5 $\mu$K arcmin, which is well above the current sensitivity of dedicated experiments. As a result, the lensing sample variance significantly contributes to the error of the measurement of *r* (Ade et al. 2021).

The CMB Stage 4 (CMB-S4) experiment is a next-generation planned ground-based experiment. Among its major science goals (Abazajian et al. 2016) is detecting PGWs at greater than 5$\sigma$, provided *r* is larger than 0.003, or putting an upper limit $r < 0.001$ at 95% C.L. in the absence of a signal (Abazajian et al. 2022), which would either confirm or reject some of the most popular inflationary models. This can be achieved through deep observations of the recombination peak of the $B$-mode power, in combination with high-resolution observations dedicated to the reconstruction of the lensing signal. These tight constraints on $r$ from degree scales can only be achieved if the lensing sample variance is removed to some fraction of its initial value (Abazajian et al. 2022). By providing a high-quality tracer of the lenses on a broad range of scales centered on lensing multipole $L \sim 500$, one can combine this with knowledge of the $E$ mode on similar scales to produce a template of the lensing $B$ mode. This is illustrated in Figure 1, where we have used the well-known perturbative, large-scale limit for the white lensing $B$-mode power (Lewis & Challinor 2006; Challinor et al. 2018),

$$C_\ell^{BB,\text{lens}} \simeq 2 \sum_L \frac{2L+1}{4\pi} C_L^{\kappa\kappa} C_L^{EE}. \quad (1)$$

---
[5] The pivot scale in this constraint is $k_\star = 0.05 \, \mathrm{Mpc}^{-1}$ and tensor spectral index $n_t = 0$.

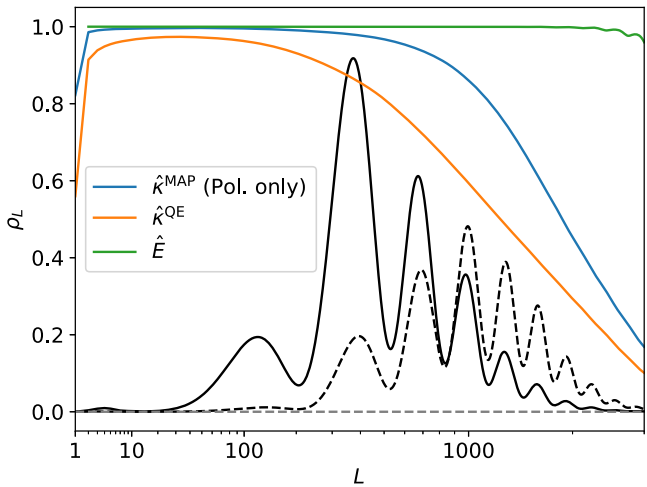

**Figure 1.** Expected fidelity of an optimal internal lensing reconstruction from idealized deep CMB-S4-like polarization data with a noise level of 0.5 $\mu$K arcmin (blue). Plotted is the expected cross-correlation coefficient, $\rho_L$, of the reconstruction to the true lensing. Quadratic estimator reconstruction is also shown for comparison (orange). The solid black line shows the contribution of lensing multipoles to the large-scale lensing $B$-mode power, and is derived from the lensing-perturbative, large-scale limit formula, Equation (1), with an arbitrary normalization. The shape of the black line shows that good lens reconstruction is relevant over the approximate range of $30 \lesssim L \lesssim 2000$. In this approximation, the contribution of the $E$-mode power to the large-scale $B$-mode power is given by the exact same curve, and the $E$ mode on the relevant scales for lensing reconstruction purposes is essentially noise-free. The expected cross-correlation coefficient to the true $E$ mode is shown in green, and ignores possible complications on large scales owing to the intricacies of ground-based observations.

Here, $C_L^{\kappa\kappa}$ is the lensing convergence power spectrum and $C_L^{EE}$ is the CMB $E$-mode power. If built correctly (Baleato Lizancos et al. 2021a), the template-delensing approach is essentially optimal when compared to complex and costly Bayesian techniques that extract all of the information (Carron 2019). As can be seen in Figure 1, the deep CMB-S4-like configuration allows for about 95% delensing, leaving a delensed kernel centered on $L \sim 1000$ (shown as the dashed line, again with an arbitrary normalization). External lensing tracers including ambitious future galaxy surveys are not expected to exceed a cross correlation of $\sim 0.7$ (Manzotti 2018) on the relevant scales, leaving little room for further improvements to internal delensing. This is in contrast to shallower experiments such as Simons Observatory (Ade et al. 2019; Namikawa et al. 2022), which can benefit from external delensing. In a practical situation, it is often the largest lenses that are hardest to reconstruct, but as can be seen, they contribute little to the $B$-mode power.

The standard tools to extract the lensing deflections from the CMB are the quadratic estimators originally designed by Hu & Okamoto (2002) and Okamoto & Hu (2003; see also the more recent study by Maniyar et al. 2021). By construction, the lensing reconstruction noise of a quadratic estimator is determined by the observed CMB total power spectra, inclusive of the lensing component and instrumental noise. While instrumental noise can be lowered with the help of more sensitive observations, the lensing component sets a fundamental limit to the precision of the estimator. However, with deep observations of the sub-degree-scale CMB polarization, where the primordial $B$ mode is negligible, a simple field-counting argument[6] suggests that the lensing deflection field can be reconstructed internally from the CMB with a high signal-to-noise ratio (S/N). Methods to overcome this limit in idealized experimental settings have long been known (Hirata & Seljak 2003a, 2003b), where usage of the CMB likelihood function is the key to improving the lensing measurement. The output is the maximum a posteriori (MAP) lensing map, conditioned on the likelihood model. More recently, this method has been revisited, cleaned of approximations, and extended to more realistic conditions (Carron & Lewis 2017), and was tested on deep POLARBEAR data (Adachi et al. 2020). Its applicability remained, however, limited to very small fractions of the sky since it used the flat-sky approximation.

As part of the CMB-S4 Collaboration planning effort (Abazajian et al. 2022), we extended the MAP lensing reconstruction method presented in Carron & Lewis (2017) to arbitrary large sky fractions and to curved sky geometry, and used it to test the performance of a map-based internal delensing forecast. The forecast starts from the component separation of simulated CMB maps all the way to the inference of $r$. The CMB-S4 simulations include inhomogeneous noise that our internal delensing algorithm takes into account accordingly. Preliminary versions of these efforts toward optimal iterative lensing reconstruction on component-separated data were recently demonstrated by Legrand & Carron (2022, 2023), focusing on the lensing auto-spectrum reconstruction. Further, this algorithm was applied to the simulated full-sky polarization maps of the Probe of Inflation and Cosmic Origins (PICO; Aurlien et al. 2023). PICO is a probe-scale space mission concept (Hanany et al. 2019) targeting $\sigma(r) \simeq 1 \times 10^{-4}$. A successful implementation of map-based internal delensing after component separation has been demonstrated, reaching the required levels of $\sigma(r)$ for most foreground models with homogeneous noise.

In this paper, we use polarization maps only. The motivation for this is simple. Although there is some additional independent information on the lensing $B$ mode that can be gleaned by including the temperature maps, when the polarization noise is low, the expected gain is small.[7] Including temperature would also substantially complicate the analysis, if only due to the expected small-scale extragalactic foregrounds (such as the cosmic infrared background, the Sunyaev–Zeldovich effect, and point-source signals), which are correlated with the lensing signal and to which standard quadratic estimators react strongly (Osborne et al. 2014; van Engelen et al. 2014; Ferraro & Hill 2018; Baleato Lizancos & Ferraro 2022; Omori 2022; Darwish et al. 2023; MacCrann et al. 2023). Nevertheless, the lensing reconstruction software, delensalot (Belkner & Carron 2023),[8] which is now publicly available, can also analyze temperature data alone or in combination with polarization data.

---

[6] In the absence of significant tensor modes, and since the deflection is almost a pure gradient, there are two unknown fields; the unlensed $E$ modes and the lensing potential, which in the low-noise limit can be reconstructed from the same number of observed fields, the lensed $E$ and $B$ (Hirata & Seljak 2003a), at least in principle.
[7] On the SPDP configuration, the main focus of this paper, we have estimated that including temperature data could reduce the residual lensing $B$-mode power by only 0.2%, provided the effects of atmospheric noise and foregrounds are perfectly under control.

This paper is structured as follows. After a general discussion dedicated to lensing reconstruction beyond the quadratic estimator, Section 2 presents the lensing reconstruction scheme. This is the most technical section of the paper, which can be omitted for readers interested in results obtained on CMB-S4 simulations. In its most general implementation, the reconstruction is optimal. We close that section with a few tests assuming full-sky coverage. Among those tests, we demonstrate in Section 2.9 a successful joint reconstruction of the lensing gradient and post-Born lensing curl deflection modes that are detectable with CMB-S4. In Section 3 we discuss the generation of the CMB-S4 simulations. Besides the lensed CMB signal, the simulations contain inhomogeneous noise. In addition, we also test different foreground models that contain non-Gaussianity signatures extending to high multipoles, both aspects potentially relevant for the lensing reconstruction. The results are presented in Section 4. We first discuss the component separation of the simulations generated in Section 3, we then present the delensing performance seen on these maps, and compare them to predictions, and finally, we feed the lensing templates to a parametric $r$-inference pipeline. We conclude in Section 5. Appendix D presents curved sky geometry calculational details relating to Section 2.

## 2. Iterative CMB Lensing Reconstruction

Intuitively, optimal lensing reconstruction works iteratively until the two following steps have converged: (i) estimation of the lensing signal with a quadratic estimator, and (ii) construction of a version of the CMB data that is delensed using that estimated lensing signal. This is what the pioneering papers on likelihood-based reconstructions effectively do (Hirata & Seljak 2003a, 2003b), albeit not very transparently.

We use a straightforward and economical approach, where gradients of a CMB lensing likelihood and prior are calculated and used to progress toward the most probable lensing map, using a variant of Newton's optimization method. However, this is certainly not the only possible way to eventually achieve efficient delensing. Several variants of *Bayesian lensing* are being developed that can potentially also be very powerful and most useful by the time CMB-S4 starts collecting data. For example, instead of finding the most probable lensing map, the CMB lensing posterior can also be probed through sampling (Anderes et al. 2015; Millea et al. 2020), as has also been demonstrated recently on data (Millea et al. 2021). In this case, keeping the computational cost under control is clearly a challenge at the moment. Reconstructing the joint most probable CMB fields and lensing map (Millea et al. 2019), instead of just the lensing map, is another possibility, which should have in principle a numerical complexity comparable to the one of this work. The lensing map obtained from the joint posterior has some unphysical features on large scales (Millea et al. 2019), but this is of moderate importance for delensing, and could potentially also be cured. Finally, the MUSE method (Millea & Seljak 2022; Bianchini & Millea 2023) is another iterative scheme that addresses the CMB lensing reconstruction problem, which could also be efficient, but was not tested here.

Before we turn to details, let us discuss, first in general terms, the analogies and differences of our method to quadratic estimation, and the physical interpretation of the CMB lensing likelihood gradients.

Crucially, for a fixed set of lenses, a deflected Gaussian field remains Gaussian, and only displays an anisotropic power spectrum, or, equivalently, an anisotropic covariance. The log-likelihood function of the lenses, $\ln p(X^{\text{dat}}|\alpha)$ (where $\alpha$ is the deflection vector field), when seen as a function of $X^{\text{dat}}$, has a Gaussian form, and therefore, always contains a term quadratic in the CMB fields. The variation with respect to $\alpha$ of this term has a form that is close to a (unnormalized) standard quadratic estimator acting on maps delensed by $\alpha$ (Hanson et al. 2010; Carron & Lewis 2017, see also later in this paper where we derive all the relevant terms in this discussion), and is sensitive to the residual lensing signal beyond $\alpha$. Further, the anisotropic covariance explicitly depends on $\alpha$. Hence, the log-likelihood gradient also contains a term independent of the data (the variation of the log-determinant of the covariance function). This is the direct analog to the *mean field* of quadratic estimation that removes signatures of anisotropy, unrelated to lensing, from the quadratic piece. Let us call these two pieces $g_\alpha^{\text{QD}}$ and $g_\alpha^{\text{MF}}$, with the total log-likelihood gradient being $g_\alpha^{\text{QD}} - g_\alpha^{\text{MF}}$, as introduced in (Carron & Lewis 2017). For $\alpha = 0$, this gradient is precisely[9] the quadratic estimator. For nonzero $\alpha$, this is a *quadratic* estimator probing residual lensing beyond $\alpha$ from $\alpha$-delensed maps. We note that the usage of quadratic here is a misuse of language since $\alpha$ itself depends on the data in a quite convoluted manner.

In a practical situation, the quadratic estimator mean field typically strongly dominates at low-lensing multipoles, but has a very red spectral profile, and therefore, becomes irrelevant for smaller-scale lenses (see Hanson et al. 2009; Benoit-Levy et al. 2013, or Planck Collaboration et al. 2020b, Figure B.1, for a detailed characterization of the components of the Planck reconstruction mean field). For perfectly idealized conditions, it is zero (because with $\alpha = 0$ and an isotropic covariance, there is no special vector field to point to). At nonzero $\alpha$, $g_\alpha^{\text{MF}}$ differs in two ways from the quadratic estimator mean field. On the one hand, the CMB maps are partially delensed by $\alpha$. This, however, does not essentially change anisotropies coming from masking or other anisotropies that leave a similar signature, but would have a different amplitude according to the changes in the data spectra. On the other hand, when producing the delensed maps, the instrumental noise is unavoidably also *delensed* by $\alpha$. This creates a new source of anisotropy in the delensed data, which was not present before delensing, and therefore, induces a mean field even for perfectly idealized conditions. To the best of our knowledge, this specific characterization of the mean field component is new to this paper and is therefore discussed in detail in Section 2.5.

---

[8] https://github.com/NextGenCMB/delensalot

[9] Up to optimization of the quadratic estimator weights to include nonperturbative effects (Hanson et al. 2011; Fabbian et al. 2019), which are irrelevant for the discussion, and up to the normalization of the estimate.

If $p(X^{\rm dat}|\boldsymbol{\alpha})$ were also Gaussian when seen as a function of $\boldsymbol{\alpha}$, then from the gradient calculated at $\boldsymbol{\alpha}=0$, a single step of Newton's method would bring us directly to the maximum; in other words, the quadratic estimator would be perfect. This is clearly not the case when the lensing effect on the total map spectra is strong, which then requires an iterative method to find the maximum point. The main technical difficulties one encounters in doing so can be split as follows:

(i) to get $g_\alpha^{\rm QD}$ we must, using the data and $\boldsymbol{\alpha}$, estimate the partially delensed CMB in the presence of masking and (possibly wildly) inhomogeneous noise, etc.;
(ii) a naive estimate of $g_\alpha^{\rm MF}$ in quadratic estimation must go through simulations, but this is too costly to perform at each and every iteration; and
(iii) all of this is costly enough numerically speaking that we must devise a scheme with fast convergence to the maximal likelihood point.

We found that converting the flat-sky tools of Carron & Lewis (2017) to the curved sky without essential modifications does deal with points (i) and (iii) to a satisfactory level: lensing reconstruction is to a large extent a localized operation in position space, and flat-sky implementations can inform in a mostly reliable manner the performance of their curved sky versions.[10] The iterative scheme we use to progress toward the maximum point is the very same as the L-BFGS scheme (Nocedal 1980) as described in Carron & Lewis (2017). Point (ii) has the potential of being more than a technical annoyance: even if a maximally efficient method to calculate an accurate mean field with just one simulation existed (which is in fact the case under some conditions, as we test later on), this would still approximately double the numerical cost. The good news is that in this paper we account for the mean field $g_\alpha^{\rm MF}$ at no additional cost compared to a standard quadratic estimator (QE) analysis, by using the QE mean field. The mean field dependence on $\boldsymbol{\alpha}$ is weak because it is almost entirely sourced by the $EE$ part of the quadratic piece. Typical mean field sources look like lensing convergence rather than shear, but large-scale convergence modes do not produce $EB$-type signatures from a pure-$E$-polarization pattern. Since the relative importance of the lensing contribution to the $EE$ CMB spectrum is much weaker than for $BB$, the change in the mean field with iterations is relatively weak. Additionally, low-lensing multipoles where the mean field is largest contribute little to degree-scale lensing $B$ modes (see Figure 1), so inaccuracies have minor impacts.

The rest of this section is dedicated to a detailed and self-contained description of our lensing reconstruction method in curved sky geometry. Section 2.1 reviews the basis of lensing remapping on the sphere, which allows us to introduce some notation as well. Section 2.2 defines more precisely our lensing likelihood model and priors. Section 2.3 obtains the lensing magnification matrix induced by the deflection on the curved sky, which our implementation requires. We then discuss $g_\alpha^{\rm QD}$ in more detail in Section 2.4, and $g_\alpha^{\rm MF}$ in Section 2.5. Finally, the construction of the lensing-induced CMB polarization templates from the lensing reconstruction outputs is discussed in Section 2.6. Throughout, we make heavy use of spin-weighted spherical harmonics transforms, and of the spin-raising and spin-lowering operators on the sphere, with the necessary mathematics reviewed in Appendix A.

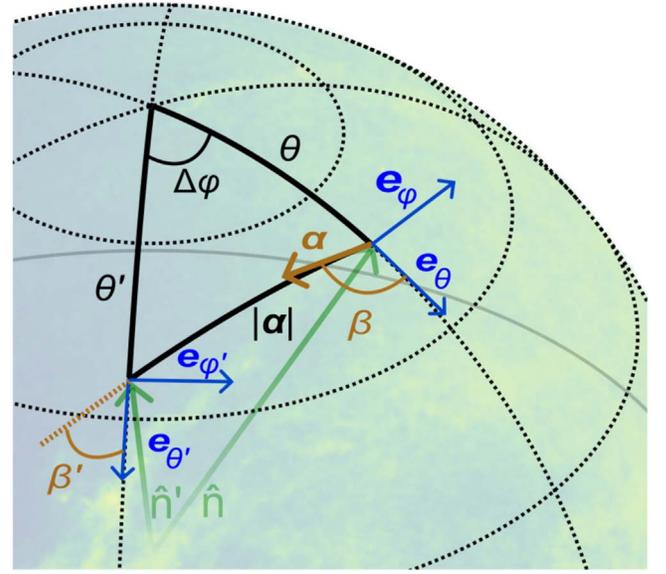

**Figure 2.** Lensing geometry on the curved sky and our notation, here for a greatly exaggerated deflection. The deflection vector $\boldsymbol{\alpha}$ lies in the plane tangent to the unit sphere at the observed position $\hat n$, pointing toward the undeflected position $\hat n'$, lying an angle $\alpha = |\boldsymbol{\alpha}|$ away on the great circle generated by $\boldsymbol{\alpha}$. The change of the local reference axes from $\hat n$ to $\hat n'$ owing to the sky curvature is characterized by the difference $\beta - \beta'$, the change in the angle between the great circle and the local basis vector $\boldsymbol{e}_\theta$ at $\hat n$ and $\hat n'$, respectively (see Equation (2)).

### 2.1. Lensing Remapping

To a good approximation,[11] the only effect of gravitational lensing on the Stokes parameters of the CMB is a remapping of points on the sphere (Lewis & Challinor 2006; Lewis et al. 2017). Within this approximation, the lensed temperature field at position $\hat n$ matches that of the unlensed one at $\hat n'$, where $\hat n'$ is defined as lying at distance $\alpha$ from $\hat n$ along the geodesic starting there in the direction $\boldsymbol{\alpha}$. Here $\boldsymbol{\alpha}(\hat n)$ is the deflection vector field, living in the space tangent to the unit sphere at $\hat n$. The geometry and angles are shown in Figure 2. For polarization or other fields with nonzero spin, the curvature of the sphere sources a typically small additional phase shift, caused by the change in the reference axes at $\hat n$ and $\hat n'$. For a spin-weighted CMB field ($_s T$), we can generally write the deflection operation ($\mathcal{D}_\alpha$) compactly as

$$[\mathcal{D}_\alpha \, _sT](\hat n) \equiv e^{is(\beta - \beta')} \, _sT(\hat n'). \qquad (2)$$

Actual implementation of the remapping on the sphere requires explicit formulae relating $\hat n'$, $\beta'$ to $\hat n$, $\beta$. These are easily gained from spherical trigonometry (Lewis 2005; Lavaux & Wandelt 2010, for example).

Let $\eth$ (or $\partial^+$) and $\bar\eth$ (or $\partial^-$) be the spin-raising and spin-lowering operators (reviewed very briefly in Appendix A). We can use the gradient and curl scalar potentials, $\phi$ and $\Omega$, to describe the two degrees of freedom of $\boldsymbol{\alpha}$, with spin-weight components,

$$_1\alpha(\hat n) = -\partial^+ \phi(\hat n) - i\partial^+ \Omega(\hat n). \qquad (3)$$

---

[10] After accounting of course for the different scaling of spherical harmonic transforms compared to Fourier transforms, etc.

[11] For degree-scale $B$ modes, leading deviations from the remapping approximation are expected to come from the combination of emission-angle and time-delay effects, relevant only for $r$ well below $10^{-5}$ (Lewis et al. 2017).

To first order and for a flat Universe,

$$\phi(\hat{n}) = -2 \int_0^{\chi_*} d\chi \left( \frac{\chi_* - \chi}{\chi \chi_*} \right) \Psi(\hat{n}, \chi), \quad (4)$$

where $\Psi$ is the Weyl potential and $\chi_*$ the comoving radial distance to the last scattering surface. There is no first-order contribution of $\Psi$ to $\Omega$ because to linear order in the scalar perturbations, the lensing curl potential $\Omega$ vanishes. We do not consider nonscalar sources here since they are expected to be small. At second order, the scalar curl has, however, some relevance at the CMB-S4 depth and is discussed further in Section 2.9. Locally, the distortions induced by $\alpha$ can also be characterized through the (scalar) convergence $\kappa$, the (scalar) field rotation $\omega$, and the spin-2 shear $_2\gamma = \gamma_1 + i\gamma_2$:

$$\kappa(\hat{n}) + i\omega(\hat{n}) = \frac{1}{2} \partial_1^- \alpha(\hat{n});$$
$$\gamma_1(\hat{n}) + i\gamma_2(\hat{n}) = \frac{1}{2} \partial_1^+ \alpha(\hat{n}). \quad (5)$$

Plugging Equation (3) into Equation (5), we see that the gradient and curl modes of the shear are uniquely set by the gradient and curl potential, respectively. It also holds that $\kappa = -\frac{1}{2}\Delta\phi$ and $\omega = -\frac{1}{2}\Delta\Omega$, where $\Delta$ is the spherical Laplacian.

Finally, for future reference, we provide the perturbative version of the lensing remapping (2), which is valid for tiny deflections (Challinor & Chon 2002):

$$[\mathcal{D}_\alpha {}_s\mathcal{T}](\hat{n}) \sim -\frac{1}{2} [_{-1}\alpha(\hat{n}) \partial_s^+ \mathcal{T}(\hat{n}) + {}_1\alpha(\hat{n}) \partial_s^- \mathcal{T}(\hat{n})]. \quad (6)$$

In $\Lambda$CDM cosmologies, however, this perturbative description is not sufficient, and Equation (2) must be applied.

### 2.2. Lensing Likelihood Model and Priors

A concrete calculation of the CMB likelihood must utilize several fiducial ingredients, which we now discuss.

We model the likelihood according to the remapping approximation, Equation (2), as follows: let $_2P^{\mathrm{unl}}$ be the unlensed polarization maps. The observed Stokes $Q$ and $U$ polarization maps are then

$$X^{\mathrm{dat}} \equiv \begin{pmatrix} Q^{\mathrm{dat}} \\ U^{\mathrm{dat}} \end{pmatrix} = \mathcal{B}\mathcal{D}_\alpha {}_2P^{\mathrm{unl}} + \mathrm{noise}, \quad (7)$$

where $\mathcal{B}$ describes the instrument beam and transfer functions, along with the projection of the lensed polarization onto its $Q$ and $U$ components. In all realistic cosmological models, the unlensed polarization is massively dominated by the $E$-mode signal. The degree scales at which the $B$-mode signal, sourced from gravitational waves, could potentially be observed, are typically excluded from the lensing reconstruction analysis (see Section 2.7). This motivates the use of a pure $E$-mode unlensed polarization as the baseline, in which case

$$_2P^{\mathrm{unl}}(\hat{n}) = -\sum_{\ell m} E^{\mathrm{unl}}_{\ell m} {}_2Y_{\ell m}(\hat{n}). \quad (8)$$

The power spectrum of this Gaussian unlensed $E$ mode, the $C_\ell^{EE,\mathrm{unl}}$ spectrum, is another fiducial ingredient of the analysis. In principle, one could attempt to jointly reconstruct this spectrum along with $\alpha$ and $E^{\mathrm{unl}}$. However, within the $\Lambda$CDM model, this spectrum is already tightly constrained (and will be even more so by the time of CMB-S4), and introducing this additional flexibility is unlikely to improve the delensing while complicating the analysis by a fair amount. Therefore, we fix it to our fiducial cosmology.

The covariance of the data conditioned on $\alpha$, $\mathrm{cov}_\alpha$, is obtained by squaring Equation (7) and averaging over the unlensed CMB, while keeping the lensing deflection fixed:

$$\mathrm{cov}_\alpha \equiv \langle X^{\mathrm{dat}} X^{\mathrm{dat},\dagger} \rangle_\alpha$$
$$= \mathcal{B}\mathcal{D}_\alpha {}_2\mathcal{Y} C^{EE,\mathrm{unl}} {}_2\mathcal{Y}^\dagger \mathcal{D}_\alpha^\dagger \mathcal{B}^\dagger + N. \quad (9)$$

Here, we have used the compact matrix times vector notation $_2P^{\mathrm{unl}} = {}_2\mathcal{Y}E^{\mathrm{unl}}$ to represent Equation (8), and assumed that the noise term is statistically independent of the signal, with noise covariance matrix $N$. The lensing likelihood then takes the form

$$-2 \ln p(X^{\mathrm{dat}}|\alpha) = X^{\mathrm{dat},\dagger} \mathrm{cov}_\alpha^{-1} X^{\mathrm{dat}} + \ln \det \mathrm{cov}_\alpha. \quad (10)$$

Of course, a likelihood that relies on several fiducial ingredients will only be approximate. For example, the noise matrix is, at least in part, only crudely known in practice. Unmodeled anisotropies in the beams or induced by the scanning strategy can also potentially introduce uncertainties in the lensing estimation. These challenges do not generally prevent the reconstruction of the lensing map, but can certainly result in a suboptimal performance of the resulting tracer. These complications are beyond the scope of this paper and are left for future work.

The lensing likelihood does not carry enough information to constrain the small-scale modes of the lensing deflection field. Supplementing the likelihood with prior information is necessary for an iterative search to converge. Including a prior is in fact not only necessary, but also desirable since the prior downweights noisy lensing modes, thereby preventing lensing reconstruction noise from adding too much $B$ power to the lensing $B$-mode template. Nonlinear effects affecting the lensing map are fairly weak on most scales relevant to delensing, making the simplest choice of a Gaussian prior on each mode both natural and close to optimal.[12] This requires introducing another fiducial ingredient, the spectrum of the lensing map. Using $\kappa$ and $\omega$ to parameterize the field, the posterior probability density function to be maximized can then be written as

$$-2 \ln p(\alpha|X^{\mathrm{dat}}) = -2 \ln p(X^{\mathrm{dat}}|\alpha)$$
$$+ \sum_{LM} \frac{|\kappa_{LM}|^2}{C_L^{\kappa\kappa,\mathrm{fid}}} + \sum_{LM} \frac{|\omega_{LM}|^2}{C_L^{\omega\omega,\mathrm{fid}}}, \quad (11)$$

where the likelihood on the right-hand side is given by Equation (10), and $C_L^{\kappa\kappa,\mathrm{fid}}$, $C_L^{\omega\omega,\mathrm{fid}}$ are the fiducial spectra. In our baseline analyses, the curl is assumed to vanish, in which case only the convergence term appears in this equation.

### 2.3. Lensing Magnification for Macroscopic Deflections

Gravitational lensing changes the solid angle under which a CMB patch is observed. In Figure 2, a lensed region of solid

---

[12] Usage of a prior ignoring the non-Gaussianity of the deflection field does not prevent reconstruction of non-Gaussian features, as long as the likelihood is constraining enough. Reconstructions on simulations with realistic, non-Gaussian input lensing maps converge just as well. A detailed study will be presented in a different paper. Using a Gaussian prior, results are insensitive to the prior spectrum shape since this is already tightly constrained from observations (Planck Collaboration et al. 2020b; Carron et al. 2022; Qu et al. 2024).

angle $d\Omega$ observed at $\hat{n}$ occupies an unlensed area $d\Omega|A(\hat{n})|$ at $\hat{n}'$, where

$$|A(\hat{n})| \equiv \left| \frac{d^2\hat{n}'(\hat{n}, \boldsymbol{\alpha}(\hat{n}))}{d^2\hat{n}} \right|, \quad (12)$$

is the determinant of the magnification matrix (the Jacobian of the sphere remapping induced by $\boldsymbol{\alpha}$). The well-known main effect is given by the convergence as $|A| \sim 1 - 2\kappa$ to first order Lewis & Challinor 2006, e.g., For this reason, delensing the CMB will slightly demagnify (for $\kappa > 0$) or magnify (for $\kappa < 0$) regions of the data, inevitably, for the purpose of reducing statistical anisotropies in the CMB signal. However, this (de)magnification induces new anisotropies in the noise map. Based on this, and for deflections larger than the coherence length of the noise, we expect the noise covariances to crudely follow the magnification,

$$N_{\text{[delensed map]}} \sim (1 - 2\kappa(\hat{n}))N_{\text{[raw map]}}. \quad (13)$$

As a result, the noise covariance becomes inhomogeneous after delensing, even if it were perfectly uniform in the data. The noise covariance decreases(increases) slightly in regions of positive(negative) convergence. This introduces a novel source of anisotropy in the iterative search for the optimal lensing map, referred to as the *delensed noise mean field*. Quadratic estimates aimed at capturing residual lensing must, in principle, subtract this *delensed noise mean field*. Further details on this effect are discussed in Section 2.5.

In our reconstruction algorithm, which is based on the likelihood function (see Equations (9) and (10)), the operator $\mathcal{D}_\alpha^\dagger$ contains the delensing operations, and its explicit form is as follows: it can be observed (using, for example, an explicit matrix representation for $\mathcal{D}_\alpha$ defined in Equation (2)), that for deflection fields that are weak enough to be invertible, the following relation holds:

$$[\mathcal{D}_\alpha^\dagger \, {}_sT](\hat{n}') = |A_{\alpha^{-1}}(\hat{n}')|[\mathcal{D}_{\alpha^{-1}} \, {}_sT](\hat{n}'). \quad (14)$$

In this equation, $\boldsymbol{\alpha}^{-1}$ is the deflection field inverse[13] to $\boldsymbol{\alpha}$, and $|A_{\alpha^{-1}}|$ is the determinant of its magnification matrix.

In practice, delensing is always applied through $\mathcal{D}_\alpha^\dagger$, on maps that are first inverse-noise-variance weighted. At the time this analysis started, the implementation of $\mathcal{D}_\alpha^\dagger$ used relation (14), and therefore, required explicit calculation of these determinants, which we now discuss.[14]

The exact form of the Jacobian determinant, and allowing for macroscopic deflections, is derived in Appendix C. The result is

$$A = \frac{\sin\alpha}{\alpha}[(1-\kappa)^2 + \omega^2 - \gamma_1^2 - \gamma_2^2]$$
$$+ \left(\cos\alpha - \frac{\sin\alpha}{\alpha}\right)(1 - \kappa - \cos(2\beta)\gamma_1 - \sin(2\beta)\gamma_2), \quad (15)$$

at each point on the sphere. The entire second line is a correction that is at least quadratic in the deflection angle $\alpha$. It tends to reduce the Jacobian for large deflections since neighboring geodesics get closer together on the sphere. Sanity checks of this formula are provided in Appendix C. Generally though, the difference in $|A|$ to the leading-order result $1 - 2\kappa$ is of course small for reasonably sized deflections. This linearization is accurate over the full sky to 0.3% at worst for a $\Lambda$CDM deflection field with $L_{\max} = 4000$. For the same configuration, neglecting the sky curvature (i.e., only considering the square bracket in Equation (15)), introduces a negligible error of $\mathcal{O}(10^{-7})$ in the determinant. This is what we used in practice.

---

[13] The deflection field $\boldsymbol{\alpha}^{-1}$ inverse to $\boldsymbol{\alpha}$ is defined as the one undoing the remapping induced by $\boldsymbol{\alpha}$, when this is possible. In the notation of Equation (2), $\mathcal{D}_{\alpha^{-1}} = [\mathcal{D}_\alpha]^{-1}$.

[14] At the time of publishing this paper, the delensing operation has been improved by Reinecke et al. (2023), and explicit calculation of the determinant is not a requirement anymore.

### 2.4. Likelihood Gradients, Quadratic Part

The gradients with respect to the lensing map of the lensing likelihood are a necessary calculation to search for the MAP deflection field. We find position space convenient for analytical work: lensing is a local effect on the CMB, and this naturally leads to compact expressions. We proceed following Appendix A of Carron & Lewis (2017), which we complete with the exact calculation of the gradients. We consider the complex variation

$$2\frac{\delta}{\delta_{\pm 1}\alpha(\hat{n})} \equiv \frac{\delta}{\delta\alpha_\theta(\hat{n})} \mp i\frac{\delta}{\delta\alpha_\varphi(\hat{n})}. \quad (16)$$

These two gradients will lower (for $\delta_1\alpha(\hat{n})$), or raise (for $\delta_{-1}\alpha(\hat{n})$) the spin of the field acted upon by one unit, respectively. We can then define the spin-1 likelihood gradient as

$$2\frac{\delta \ln p(X^{\text{dat}}|\alpha)}{\delta_{-1}\alpha(\hat{n})} \equiv {}_1g_\alpha^{\text{QD}}(\hat{n}) - {}_1g_\alpha^{\text{MF}}(\hat{n}), \quad (17)$$

where we have split the total likelihood gradient into its contribution from the quadratic part ($g_\alpha^{\text{QD}}$), and from the log-determinant ($-g_\alpha^{\text{MF}}$). The decomposition of ${}_1g_\alpha(\hat{n})$ into gradient and curl components gives the lensing potential and curl potential gradients.

It is apparent that the relevant variations entering Equation (17) are those of deflected spin-weighted fields $\mathcal{D}_\alpha \, {}_sT$ under a change in the deflection vector. On the flat sky, where the remapping equation at position $\hat{n}$ is merely $[\mathcal{D}_\alpha \, {}_sT](\hat{n}) = {}_sT(\hat{n} + \boldsymbol{\alpha})$, the gradient with respect to $\boldsymbol{\alpha}$ is simply the gradient of the field evaluated at the deflected position. On the curved sky, we expect a similar-looking result, with (when working with the spin-weighted components) the standard gradient replaced by minus the spin-raising operator. If the deflection is large enough that the sky curvature is relevant between $\hat{n}$ and $\hat{n}'$, there will be possible corrections. As we now discuss, it turns out that they can be calculated exactly, but are also safely negligible.

The concrete calculation is deferred to Appendix D. The result is

$$2\frac{\delta[\mathcal{D}_\alpha \, {}_sT](\hat{n})}{\delta_{\pm 1}\alpha(\hat{n})} = -[\mathcal{D}_\alpha \, \partial_s^\mp T](\hat{n})$$
$$+ f(\alpha(\hat{n}))([\mathcal{D}_\alpha \partial_s^\mp T](\hat{n}) - e^{\mp 2i\beta}[\mathcal{D}_\alpha \partial_s^\pm T](\hat{n})). \quad (18)$$

Here, ${}_sT$ is the corresponding spin-$s$ field, and we have omitted its spin subindex on the right-hand side of the equation. The first line is the direct analog of the flat-sky result, and the second line is the correction term. The prefactor $f(\alpha)$ is $\frac{1}{2}\left(1 - \frac{\sin\alpha}{\alpha}\right) \sim \alpha^2/12$ for small $\alpha$. The second line accounts for the focusing effect of the spherical geometry on nearby geodesics. For example, all geodesics of length $\pi$ starting from a point meet again at the antipodal point. This provides a useful sanity check of this formula, performed in Appendix D, which is that any variation of $\alpha$ at $\hat{n}$ that leaves the deflection angle $\alpha(\hat{n})$ invariant and equal to $\pi$ must result in a vanishing gradient (Equation (18)). For any realistically sized $\Lambda$CDM CMB weak-lensing deflection field of a couple of arcmin, $f(\alpha) \simeq 5 \times 10^{-8}$, hence is a negligible correction. There are no practical difficulties in including this correction term and we tested its impact for both temperature and polarization estimators and on the full sky and found that it has a negligible impact: the relative improvement on the cross-correlation coefficient improves only marginally.

Safely neglecting this correction, the precise expression for the likelihood gradient takes a form perfectly analogous to the case of the quadratic estimator. As is well known (Okamoto & Hu 2003; Lewis & Challinor 2006; Maniyar et al. 2021), the quadratic estimator constructs an estimate of the lensed CMB from the data, and weights its gradient in real space by inverse-variance-weighted residuals. In our context, the only difference is that the lensing likelihood uses the current estimate of lensing map $\alpha$ as an additional piece of knowledge. The calculation of the gradient can be described as follows:

In a first step, we estimate $\alpha$ using a standard QE, and construct an estimate of the unlensed CMB, referred to as $E^{\mathrm{WF}}_\alpha$. This is in contrast to the QE for which the lensed CMB is used. We call this CMB reconstruction step *Wiener filtering* since it has the precise meaning of reconstructing the MAP point of the assumed Gaussian $E^{\mathrm{unl}}$ signal, conditioned on $\alpha$ and the other likelihood ingredients[15] being the truth. Its explicit form is

$$E^{\mathrm{WF}}_\alpha = [C^{EE,\mathrm{unl},-1} + N_\alpha^{-1}]^{-1}\, _2\mathcal{Y}^\dagger \mathcal{D}^\dagger_\alpha \mathcal{B}^\dagger N^{-1} X^{\mathrm{dat}}, \quad (19)$$

where $N_\alpha$ is the delensed $E$-noise covariance matrix,

$$N_\alpha^{-1} \equiv {}_2\mathcal{Y}^\dagger \mathcal{D}^\dagger_\alpha \mathcal{B}^\dagger N^{-1} \mathcal{B} \mathcal{D}_\alpha\, _2\mathcal{Y}. \quad (20)$$

In practice, the large bracketed matrix inverse is performed with a conjugate-gradient solver. In the second step, we construct the inverse-variance-weighted polarization residuals, which we write as $_2\bar{P}_\alpha(\hat{n})$. The residuals are the difference between the data maps ($X^{\mathrm{dat}}$) and the prediction of the likelihood model assuming that the deflection and Wiener-filtered CMB are the truth. This conditioned prediction is the beamed and deflected polarization $\mathcal{B}\mathcal{D}_\alpha P^{\mathrm{WF}}_\alpha$, where the Wiener-filtered Stokes polarization $P^{\mathrm{WF}}_\alpha$ is obtained from its $E$ mode $E^{\mathrm{WF}}_\alpha$ in the usual way, Equation (8). Inverse-variance weighting the residuals gives

$$_2\bar{P}_\alpha(\hat{n}) \equiv [\mathcal{B}^\dagger N^{-1}(X^{\mathrm{dat}} - \mathcal{B}\mathcal{D}_\alpha P^{\mathrm{WF}}_\alpha)](\hat{n}). \quad (21)$$

With this in hand, the quadratic piece of the likelihood gradient is

$$_1g^{\mathrm{QD}}_\alpha(\hat{n}) = -\sum_{s=\pm 2} {}_{-s}\bar{P}_\alpha(\hat{n})[\mathcal{D}_\alpha \partial^+_s P^{\mathrm{WF}}_\alpha](\hat{n}), \quad (22)$$

where, for the reasons discussed above, we have only included the first line of Equation (18) in the brackets. This equation is the exact analog of the standard unnormalized, quadratic estimators built from polarization (Hu & Okamoto 2002), when written in the spin-weight formalism (Planck Collaboration et al. 2020b). Since $\bar{P}_\alpha$ contains both $E$ and $B$ components, this gradient is the combination of an $EE$ and an $EB$ quadratic piece, with the $EB$ piece containing most of the signal. There is no $BB$ piece since our fiducial model assumes vanishing primordial $B$ modes.

### 2.5. Likelihood Gradient, Mean Field

Our goal in this section is to quantify the importance of the *delensed noise* mean field, which we introduced briefly in Section 2.3. We will demonstrate that it plays a very minor role in the configurations of interest in this paper.

The mean field is the part of the likelihood gradient that originates from the determinant term in Equation (10). As is the case with standard quadratic estimator analyses, its role is to remove signatures of anisotropies picked up by the quadratic piece that are unrelated to the lensing signal of interest: this can be seen from the identity

$$_1g^{\mathrm{MF}}_\alpha(\hat{n}) = \langle {}_1g^{\mathrm{QD}}_\alpha(\hat{n})\rangle, \quad (23)$$

where the average is over realizations of the data according to the likelihood model $p(X^{\mathrm{dat}}|\alpha)$, with the deflection field set to $\alpha$ in each of these realizations.[16] In the case of the quadratic estimator (for which $\alpha = 0$ in Equation (23)), and in the absence of nonidealities like masking or noise inhomogeneities, the mean field vanishes. For the iterative process, this is not true anymore because $\alpha$, the current best estimate of the lensing field, sources itself a mean field: as discussed in Section 2.3, by moving the data slightly around, it effectively magnifies or demagnifies certain areas, and the delensed noise on the $E$ mode becomes inhomogeneous even if it was homogeneous before.

We calculate and discuss this contribution in the absence of other sources of anisotropies in Appendix B. For high-resolution experiments, with polarization noise spectrum $N_\ell$, there is the simple result

$$g^{\mathrm{MF},\kappa}_{LM} \simeq -2\kappa_{LM} \sum_\ell \frac{2\ell+1}{4\pi}\left(\frac{C^{EE,\mathrm{unl}}_\ell}{C^{EE,\mathrm{unl}}_\ell + N_\ell}\right). \quad (24)$$

This is valid for low lensing $L$s, but does not fail too badly at high $L$s either. The sign is negative because a magnified ($\kappa > 0$) CMB patch will have its noise map appear demagnified after delensing. The curl component $g^{\mathrm{MF},\omega}$ is much smaller since the lensing curl potential affects the magnification to second order.

Equation (24) comes from the $EE$-only part of the iterative quadratic estimator. Although the $EB$ estimator captures much more signal, its contribution to the convergence mean field is small. This can be understood as follows: a large converging lens does not locally change the pure-$E$ property of the unlensed $E$ modes. Hence, the $EB$ part does not generally respond to large-scale convergence-like anisotropies.

To assess the importance of this mean field component, it is instructive to compare its magnitude to that of the other gradients. Doing so allows us now to provide the analytical argument that this mean field component is small enough that it can only play a role when the MAP iterative search has already converged to very good accuracy, but is negligible on the onset of the iterations.

---

[15] Explicitly, these ingredients are the inverse-noise matrix, the transfer function, and the unlensed $E$-mode power spectrum.

[16] The above equation follows from the fact that under any parameter variation $\delta$ we must have $0 = \delta\langle 1\rangle = \langle \delta \ln p(X^{\mathrm{dat}}|\alpha)\rangle = \langle g^{\mathrm{QD}}\rangle - \langle g^{\mathrm{MF}}\rangle = \langle g^{\mathrm{QD}}\rangle - g^{\mathrm{MF}}$. The last equality holds since the gradient of the log-determinant is data independent.

For deep experiments like CMB-S4, the number of noise-free $E$ modes is high. For a beam with full width at half-maximum (FWHM) of $1'$ and polarization noise of $0.5\,\mu$K arcmin, we find from Equation (24),

$$g_{\rm LM}^{\rm MF,\kappa} \simeq (-3 \cdot 10^6)\kappa_{\rm LM}. \quad (25)$$

If the mean field component is neglected, then a fully converged iterative solution must have the prior and quadratic pieces in equilibrium, that is,

$$g_{\rm LM}^{\rm PR,\kappa} = -\frac{\kappa_{\rm LM}}{C_L^{\kappa\kappa,\rm fid}} = -g_{\rm LM}^{\rm QD,\kappa}. \quad (26)$$

At the peak of the lensing power of $L \sim 30$, we find a similar value:

$$-\frac{\kappa_{\rm LM}}{C_{L=30}^{\kappa\kappa,\rm fid}} \simeq (-4.5 \cdot 10^6)\kappa_{\rm LM}. \quad (27)$$

On the other hand, in the first few iterations, the quadratic piece still completely dominates. The first iteration gradient is the (unnormalized) standard quadratic estimator, $g_{\rm LM}^{\rm QE,\kappa}$. The standard $EB$ quadratic estimator response in the large lens limit is

$$g_{\rm LM}^{\rm QE,\kappa} \simeq 2\kappa_{\rm LM}\sum_\ell \frac{2\ell+1}{4\pi}\left(\frac{C_\ell^{EE,\rm lens}}{C_\ell^{EE,\rm lens}+N_\ell}\right)\left(\frac{C_\ell^{EE,\rm lens}}{C_\ell^{BB,\rm lens}+N_\ell}\right). \quad (28)$$

This equation holds for $L \geqslant 2$.[17] Since the quadratic estimator already resolves the spectrum at low $L$ with good S/N, we can directly compare the result to Equation (25), finding

$$g_{\rm LM}^{\rm QE,\kappa} \simeq (1.7 \cdot 10^8)\kappa_{\rm LM}, \quad (29)$$

for the same, deep configuration. This suggests, that at low $L$, the delensed noise mean field piece is small. Only if the MAP method reduces the difference between the true and estimated convergence by a factor of about 70 compared to the QE estimate, does this mean field become a bias on the reconstruction which may be taken into account. At higher $L$, it is irrelevant in any case. In practice, the mean field is heavily dominated by masking, as well as other instrumental nonidealities that are either poorly known or more difficult to include in the likelihood model for practical reasons. Most relevant contributions of the mean field peak at low $L$, but they contribute little to $B$-mode delensing. This motivates a simple baseline solution as follows (and comes at zero additional cost): we perform the iterations using a constant mean field, calculated, for example, from quadratic estimators on a set of simulations with accurate noise maps and varying lensing fields. If necessary, after the iterative scheme has converged to an approximate MAP solution, $\hat{\alpha}^{\rm MAP}$, we can correct for residual mean field contamination using

$$\hat{\alpha}^{\rm MAP} - \langle\hat{\alpha}^{\rm MAP}\rangle_{\rm MC}, \quad (30)$$

---

[17] The polarization estimator dipole response is much weaker since a lensing dipole does not produce any shear.

instead of $\hat{\alpha}^{\rm MAP}$ for the delensing step, where the average is built from the set of simulations for which the iterative solutions are constructed. Since these simulations have varying input lensing, Equation (30) does not contain the $\alpha$-induced part, which is neglected altogether in this approach. Sections 2.8 and 4 contain explicit tests of this and of other options, confirming that the mean field is not a major issue for the purpose of $B$-mode delensing.

### 2.6. Construction of the B-polarization Lensing Templates

We now discuss our construction of the lensing $B$-mode template from the reconstructed lensing and $E$-mode tracers.

At the end of the iterative process, we can calculate the prediction of the lensing-induced CMB polarization as follows. The Wiener-filtered $E$ mode from Equation (19) (with, in that equation, $\alpha = \alpha^{\rm MAP}$) is our best estimate of the unlensed $E$ mode, which we write as $\hat{E}^{\rm unl}$. We first build its Stokes polarization maps, $_2\hat{P}^{\rm unl}$, and then remap it,

$$_2\hat{P}^{\rm lens}(\hat{n}) = [\mathcal{D}\hat{\alpha}^{\rm MAP}\,_2\hat{P}^{\rm unl}](\hat{n}), \quad (31)$$

where

$$_2\hat{P}^{\rm unl}(\hat{n}) = -\sum_{\ell m}\hat{E}_{\ell m}^{\rm unl}\,_2Y_{\ell m}(\hat{n}). \quad (32)$$

The $B$ mode of the deflected polarization $_2\hat{P}^{\rm lens}$ of Equation (31) is the $B$-mode lensing template, $\hat{B}^{\rm LT,MAP}$. In several parts of this work, we will use the following shorter notation for these operations:

$$\hat{B}^{\rm LT,MAP} \equiv \hat{\alpha}^{\rm MAP}\circ\hat{E}^{\rm unl}. \quad (33)$$

We will often compare the performance of iterative delensing to that of quadratic estimator delensing. One possibility is to proceed in the same manner,

$$\hat{B}^{\rm LT,QE} \equiv \hat{\alpha}^{\rm QE}\circ\hat{E}^{\rm lens}, \quad (34)$$

where the $E$-mode map $\hat{E}^{\rm lens}$ is the Wiener-filtered lensed CMB. There is a known subtlety, though, that when combining the *lensed* $E$ mode with the quadratic estimator, it is more powerful to apply the remapping to leading order in the deflection, rather than exactly (Baleato Lizancos et al. 2021a). This is due to a beneficial cancellation of nonperturbative effects, which does not occur anymore when the exact remapping Equation (31) is used on the lensed $E$ mode. Hence, for quadratic estimator delensing, we also consider the perturbative action of lensing (given by Equation (6)) on the $E$ mode.

There is less flexibility in the choice of the $E$ template in the case of the iterative lensing solution: by design of the lensing reconstruction process, the reconstruction noise in $\hat{\alpha}^{\rm MAP}$ and $\hat{E}^{\rm unl}$ have a strong statistical dependence: when the CMB likelihood model is inaccurate, both $\hat{E}^{\rm unl}$ and $\hat{\alpha}^{\rm MAP}$ can be somewhat wrong, but still manage to provide a reliable $B$-mode template. In this case, independent adjustments to $\hat{E}^{\rm unl}$ or $\hat{\alpha}^{\rm MAP}$ can break this dependence and degrade the delensing performance. The construction described by Equation (31) consistently gives the best results.

### 2.7. Low-ℓ B-mode Deprojection

The search for primordial $B$-mode power occurs on degree scales, with the help of the lensing template built from higher-resolution data. In principle, it is possible to include the very same degree-scale modes from the high-resolution data to build the lensing template. However, this is generally not advisable as these shared CMB modes induce a statistical dependence between the lensing tracer reconstruction noise and the map that is to be delensed (Teng et al. 2011; Carron et al. 2017; Namikawa 2017; Baleato Lizancos et al. 2021b). At the QE level, this appears notably as a very strong disconnected four-point function, given by the Gaussian pairings of

$$\hat{C}_\ell^{B^{\rm dat}B^{\rm LT,QE}} \ni B^{\rm dat} \cdot \hat{\alpha}^{\rm QE}(E^{\rm dat}, B^{\rm dat}) \circ \hat{E}^{\rm lens}, \quad (35)$$

with a tendency to decrease both the $B$-mode power and its variance; however, it is unrelated to true delensing. At low-noise levels, a six-point function also becomes relevant, as studied in some detail in Namikawa (2017) and Baleato Lizancos et al. (2021b), and predictions of the spectra are rendered much more complicated, particularly in the presence of tensor modes for which $B^{\rm unl} \neq 0$.

Less is known in the case of the iterative estimator, but this bias is even larger there (as we show later in Section 2.8). The lensing reconstruction gives little weight to these large-scale $B$ modes. Hence, a simple solution that we adopt here is to exclude these overlapping modes from the beginning. We proceed here in a quite careful manner with the curved sky version of the *overlapping B-mode deprojection* (OBD) technique introduced in Adachi et al. (2020). We exclude modes as follows: let $\mathcal{P}_{\ell m}(\hat{n}_i)$ be the Stokes polarization pattern on pixel $\hat{n}_i$ produced by a CMB sky $B_{\ell m}$ mode. We assign a high noise level to a set of these patterns by augmenting the noise matrix in the following way, $N \to N + \mathcal{P}^\dagger \sigma^2 \mathcal{P}$, where $\sigma^2$ is a diagonal matrix of dimension the number of patterns considered, with values on the diagonal the assigned noise variances of the patterns. According to the Woodbury matrix identity (Hager 1989), this has the effect of replacing the inverse-noise matrix model $N^{-1}$ by

$$N^{-1} - N^{-1}\mathcal{P}^\dagger \left[\frac{1}{\sigma^2} + \mathcal{P}N^{-1}\mathcal{P}^\dagger\right]^{-1} \mathcal{P}N^{-1}. \quad (36)$$

In this way, all patterns included in $\mathcal{P}$ are perfectly masked from the analysis in the limit of infinite $\sigma^2$. A high but finite value of $\sigma^2$ is often required for practical reasons but works equally well. We deproject all modes $B_{\ell m}$ with $2 \leqslant \ell \leqslant 200$, for a tiny loss of S/N,[18] using $\sigma^2 = (10^3 \mu K)^2$. The matrix inside the square brackets in this expression is therefore a dense matrix with about $\ell_B^4$ elements. The matrix is precomputed and stored in memory during the reconstruction. An alternative to this computationally rather expensive calculation of the OBD matrix would be to filter the data maps by setting the transfer function for these modes to zero. On the full sky with homogeneous noise, the two procedures are exactly equivalent. On the masked sky, it is less straightforward to cleanly exclude this set of sky modes, potentially sourcing residual internal delensing biases. We have not pursued the alternative approach in this work and leave it to a future study.

### 2.8. Idealized Full-sky Delensing

As a first step, we demonstrate the effectiveness of our lensing reconstruction method and of the $B$-mode delensing procedure on the full sky. We generate full-sky curved sky-lensed CMB data with no primordial $B$ modes, and use different $\phi$ and $E$ modes for each realization, synthesized from a standard $\Lambda$CDM cosmology. The simulations use an idealized, low-noise, high-resolution configuration: we add an isotropic Gaussian noise of $\Delta_{\rm P} = 0.5 \, \mu$K arcmin and convolve the maps with a beam FWHM of $1'$. We reconstruct the lensing potential from the $E$ and $B$ maps using both the QE and MAP methods. We restrict to the multipole range $\ell \in [2, 3000]$ for the $E$ modes and $\ell \in [200, 3000]$ for the $B$ modes to avoid internal delensing bias, as discussed in Section 2.7.

For the QE $B$-mode template estimation, we compare two approaches, as described in Section 2.6. The first approach is a perturbative remapping of the Wiener-filtered and lensed $E$ mode at first order, following Equation (6). The deflection field is taken as the Wiener-filtered QE estimate,

$$\kappa_L^{\rm WF} = \hat{\kappa}_L^{\rm QE} \frac{C_L^{\kappa\kappa}}{C_L^{\kappa\kappa} + N_L^{(0),\kappa{\rm QE}}}, \quad (37)$$

with $N_L^{(0),\kappa{\rm QE}}$ the dominant term in the variance of the reconstructed QE power spectrum, arising from the Gaussian contractions of the four-point function of the lensed CMB fields.[19] The second $B$-mode template is estimated with the nonperturbative lensing remapping of Equation (31), where the unlensed polarization field is given by Equation (19), and the deflection field is the QE field. In practice, this corresponds to the first iteration of the MAP algorithm. We found that both approaches give a similar residual $B$-mode lensing power. As shown in Baleato Lizancos et al. (2021a), the perturbative remapping of the observed (Wiener-filtered) polarization field has a similar residual $B$-mode lensing power as the nonperturbative remapping of the unlensed polarization.

For the MAP approach, the $B$-mode template is derived using Equation (31), hence, by remapping the unlensed polarization field by the MAP lensing estimate.[20] As discussed in Section 2.5, the QE mean field is zero, but the MAP mean field is not. We estimate the mean field at first order in the lensing field estimate, as described in Appendix B of Carron & Lewis (2017). Figure 3 shows the power spectrum of the lensing $B$ mode (orange), and residual lensing $B$ modes after the subtraction of the QE (dark blue), and MAP (light blue) $B$-mode templates. The QE method removes significant power of the lensed $B$ modes, thereby decreasing the mean $C_\ell^{BB}$ amplitude in the range of $\ell \in [2, 200]$ from $2.1 \times 10^{-6}$ to $4.8 \times 10^{-7} \, \mu\rm K^2$. The MAP method reduces it further, leaving about $1.3 \times 10^{-7} \, \mu\rm K^2$ of the $B$ power. For illustration purposes, we also estimate the MAP $B$-mode template for which the $B$ modes $\ell \leqslant 200$ are included in the lensing reconstruction (solid gray), and additionally show the power spectrum of the simulation's white noise (dashed gray). Including these modes results in an internal delensing bias (see Equation (35)), as can clearly be seen: the residual $B$-mode power is below the noise level of the polarization maps. In the standard analysis, where we reconstruct the lensing potential from the $B$ modes with $\ell \in [200, 3000]$, this internal delensing bias appears at scales $\ell > 200$ for both the QE and the MAP methods, albeit it is much smaller in the former case, as we can see in Figure 3.

---

[18] In the CMB-S4 South Pole deep configuration of this paper, deprojection of these modes causes the predicted residual $B$-power amplitude after delensing to rise formally to 0.053 from 0.052.

[19] We find that including $N_L^{(1),\kappa{\rm QE}}$ to the Wiener filter reduces the residual lensing amplitude by about 2%.

[20] We find that using the perturbative remapping, together with the lensed $E$ map, results in an almost identical MAP residual lensing $B$-mode spectrum, in agreement with the findings discussed in Baleato Lizancos et al. (2021a).

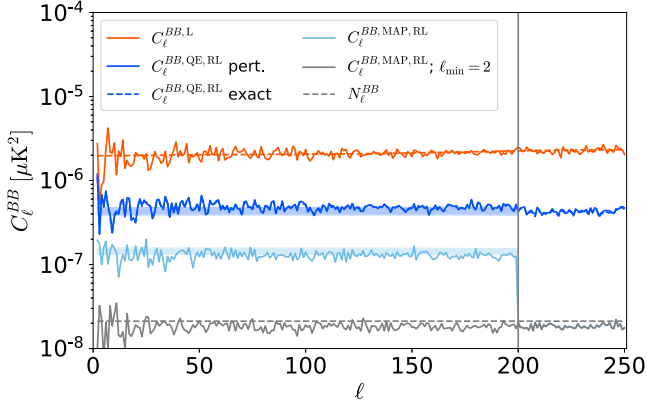

**Figure 3.** Residual $B$-mode power after subtracting the $B$ template estimated with the QE (blue) or with the MAP (light blue), in an idealized full-sky reconstruction, on the scales relevant for inference on $r$. The blue-dashed line overlaps with the solid blue and is therefore invisible; it corresponds to using the lensing remapping of the unlensed $E$ obtained at the first iteration of the MAP and it matches almost exactly the perturbative remapping of the lensed $E$ modes. Colored bands show the corresponding fiducially delensed CMB spectra (see the discussion in the text). The orange line denotes the $B$ power of the map before template subtraction (including noise) and the orange-dashed line denotes the fiducial lensed $B$ power including noise. The gray-dashed line shows the instrumental polarization noise power. To calculate the light blue line, the $B$ modes below $\ell = 200$ are not used to estimate the $B$ templates. This avoids internal delensing bias; the result of not treating internal delensing bias is illustrated by the gray line, in which case the $B$ template is estimated using all modes from $\ell = 2$. Interpreted naively, the line shows unrealistically efficient delensing with a delensed $B$-mode power slightly below the instrumental noise power of the map. The light blue line follows the gray line for $\ell \geqslant 200$.

We can compare our results with the prediction of the residual lensing $B$-mode power. To build predictions, we generally follow the procedure detailed in Legrand & Carron (2022), a variant on the original procedure for the $EB$ estimator put forward by Smith et al. (2007). The MAP solution is analytically intractable even under idealized conditions, and this simplified procedure is not free from some ambiguities, as we now discuss.

Predictions are estimated by computing the delensed power spectrum $C_\ell^{BB,\mathrm{delens}}$ from the unlensed $E$ spectrum $C_\ell^{EE,\mathrm{unl}}$ and a lensing power spectrum $C_L^{\kappa\kappa,\mathrm{lens}}$ reduced by some fraction:

$$C_L^{\kappa\kappa,\mathrm{lens}} = (1 - \epsilon_L)\, C_L^{\kappa\kappa}, \quad (38)$$

where the delensing efficiency $\epsilon_L$ is

$$\epsilon_L = \frac{C_L^{\kappa\kappa}}{C_L^{\kappa\kappa} + N_L^{\kappa\kappa}}. \quad (39)$$

This variance of the reconstructed lensing potential $N_L^{\kappa\kappa}$ contains the Gaussian contribution $N_L^{(0)}$, as well as terms higher-order in $C_L^{\kappa\kappa}$. For the QE, we found that the $N_L^{(1)}$ term (the secondary connected contractions at first order in $C_L^{\kappa\kappa}$) has an impact of roughly 20% on the amplitude of the predicted residual $B$ mode, from $3.9 \times 10^{-7}\ \mu K^2$ when $N_L^{\kappa\kappa} = N_L^{(0)}$ to $4.8 \times 10^{-7}\ \mu K^2$ when $N_L^{\kappa\kappa} = N_L^{(0)} + N_L^{(1)}$. The case including $N_L^{(1)}$ provides a better fit to the observed delensed power. For the MAP prediction, we compute these lensing spectrum biases iteratively following Legrand & Carron (2022): starting with the noise of the QE reconstruction, $N_L^{\kappa\kappa}$, we compute the delensed spectrum of Equation (38), and the corresponding partially lensed $E$ power spectrum. These spectra are then inserted as the weights and lensing response in the analytical expression of $N_L^{\kappa\kappa}$. Finally, we obtain a new delensing efficiency $\epsilon_L$ and repeat the calculation with an iterative procedure until convergence. Contrary to Smith et al. (2012), we do not include the imperfect knowledge of the $E$-mode power in the iteration procedure. We found that it does not impact the predictions, due to the low-noise level in polarization considered here. It appears that the prediction of the MAP delensing residual provides a better fit when we only include the $N_L^{(0)}$ bias in the iterations. By including the $N_L^{(1)}$ bias in all iteration steps we obtain a mean $B$-mode amplitude of $1.6 \times 10^{-7}\ \mu K^2$ across the relevant scales. However, not including it gives a mean $B$-mode amplitude of $1.3 \times 10^{-7}\ \mu K^2$, which matches the observed delensing residual from our simulation. Nevertheless, we do not have an analytical argument for whether or not the $N_L^{(1)}$ bias should be taken into account for the lensing power spectrum noise of the MAP algorithm. We take this as a measure of our uncertainty in the prediction, and show the predicted delensing amplitudes as shaded areas in Figure 3. The lower limit corresponds to $N_L^{\kappa\kappa} = N_L^{(0)}$, while the upper limit corresponds to $N_L^{\kappa\kappa} = N_L^{(0)} + N_L^{(1)}$.

The reconstruction described above used a perturbative approximation to the delensed noise mean field, which is the only source of the mean field in this idealized reconstruction. We now test the importance of this term explicitly. Figure 4 shows the spectrum of the quadratic piece $g_\alpha^{\mathrm{QD}}$ of the gradients at the starting point (the quadratic estimator, in blue) and at iteration 15 in orange, where we often stop the reconstruction for CMB-S4-like configurations. We show results for $L \geqslant 2$ because the lensing dipole behaves differently.[21] The displayed spectra are the gradients normalized by our prediction of the iterative estimator response, $\mathcal{R}_L^{\mathrm{MAP}}$,

$$\frac{g_{\alpha,LM}^{\mathrm{QD}}}{\mathcal{R}_L^{\mathrm{MAP}}}, \quad (40)$$

such that, very crudely, the normalized map corresponds to the expected Newton increment to the lensing map whenever the prior is irrelevant. The prior term is shown in purple. The two gradients are mostly in equilibrium at $L \gtrsim 2000$.[22] As can be seen in Figure 4, for $L \lesssim 1000$, the prior is substantially smaller, which suggests that there is still some amount of information in the likelihood that can be collected, although it is small.

The estimated mean field for iteration step 15 is shown in green. To compute it, we used a trick presented in Appendix 2 of Carron & Lewis (2017): instead of $\langle g_\alpha^{\mathrm{QD}} \rangle$, we calculate the average of a similar map,

$$_1\hat{g}_\alpha^{MF}(\hat{n}) = -\sum_{s=\pm 2} \langle\, _{-s}\bar{p}_\alpha(\hat{n})[\mathcal{D}_\alpha \partial_s^+ p_\alpha^{WF}](\hat{n})\rangle, \quad (41)$$

where $\bar{p}_\alpha = \mathcal{B}^\dagger x$, $p_\alpha^{WF} = C^{EE,\text{unl}}\mathcal{D}_\alpha^\dagger \mathcal{B}^\dagger \text{Cov}_\alpha^{-1} x$, $x$ is a map of unit variance, independent Gaussian variables, and $s$ is the spin of the field. Assuming the CMB likelihood model correctly describes the data, the average, Equation (41), is the same as $\langle g_\alpha^{QD}\rangle$, but requires orders of magnitude less Monte Carlo simulations for the same precision of the result. This is because the $N^{(0)}$ noise of this quadratic estimator is much lower than that of the one with the correct weights: see Appendix 2 of Carron & Lewis (2017) for more details. As derived in Section 2.5, the mean field shape follows that of the convergence (shown in black). At the lowest multipoles, it appears negative, but this may be attributed to residual Monte Carlo noise in that range. The mean field remains much smaller than the quadratic gradient piece. Only at iteration 35 (red) does the latter reach an amplitude comparable to the former. Between iterations 15 and 35, the delensing efficiency has hardly grown, from 94.5% to 95.0%.

---

[21] On the one hand, the lensing dipole has much higher noise than for $L \geqslant 2$ because it is insensitive to the shear part of the signal. On the other hand, the dipole receives a strong contribution from the aberration of the CMB caused by our motion relative to the CMB frame (Planck Collaboration et al. 2014).
[22] In this reconstruction, the size of the Newton step in the iterative procedure was suppressed for $L \geqslant 1000$ compared to smaller $L$'s, in order to avoid too large steps. This causes the feature seen at $L \sim 1000$ in the orange line.

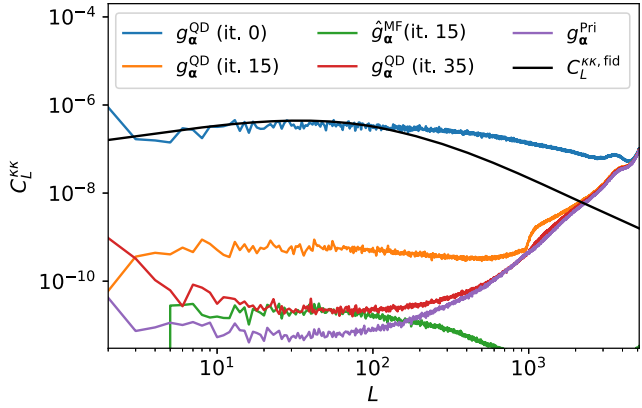

**Figure 4.** Comparison of normalized CMB likelihood gradients for a full-sky idealized reconstruction. The blue and orange curves show the quadratic piece spectrum at iteration 0 (the quadratic estimator) and 15, where we often stop the iterations. The prior term at the same iteration is shown in purple. The delensed noise mean field term, shown in green and obtained with the steps given in the text, is very small and has negligible impact on the quality of the reconstruction. We also show the quadratic piece after 35 iterations in red. Only then does it become comparable to the mean field on large scales. As discussed in the text, the improvement achieved in going from 15 to 35 iterations is tiny, however.

Finally, using Equation (41), it is possible to run the entire full-sky reconstruction process using a single Monte Carlo simulation to obtain the mean field. As expected from Figure 4, we find no significant difference. All of this demonstrates that in general the delensed noise part of the mean field can be safely neglected in the reconstruction from polarization data.

### 2.9. Idealized Joint $\phi$-$\Omega$ Lensing Reconstruction

Even though the lensing curl vanishes to first order for density perturbations,[23] it is well known that the combined effect of two shearing lenses (with misaligned shears) does induce field rotation. The leading contribution to quadratic order can be written as

$$\omega(\hat{n}) = -4\int_0^{\chi_*} d\chi \left(\frac{\chi_* - \chi}{\chi\chi_*}\right)\int_0^\chi d\chi' \left(\frac{\chi - \chi'}{\chi'\chi}\right)$$
$$\cdot [\gamma_1(\hat{n},\chi)\gamma_2(\hat{n},\chi') - \gamma_2(\hat{n},\chi)\gamma_1(\hat{n},\chi')], \quad (42)$$

and is referred to as *lens–lens coupling* (see Hirata & Seljak 2003a) or Pratten & Lewis 2016 for recent discussions). Here, $_2\gamma(\hat{n},\chi)$ is the spin-2 shear as defined in the usual way (see Equation (5)), but acting on the Weyl potential along the ray. At the noise levels of CMB-S4, this lensing curl mode can be detected in cross correlation to external data (Robertson & Lewis 2023). In this section, we test the impact of the expected curl signal by performing a joint reconstruction of the lensing gradient and curl modes. This will allow us to confirm that the lensing curl is not a source of worry for CMB-S4 delensing. Although this joint analysis doubles the number of modes to be reconstructed simultaneously, we note that this is not expected to significantly degrade the quality of the gradient reconstruction: in the absence of a parity-violating $C_\ell^{EB}$ or $C_\ell^{TB}$ signal, the lensing gradient and curl CMB quadratic reconstruction share no leading Gaussian covariance[24] $N^{(0)}$ (Namikawa et al. 2012). It is true though that each component (here convergence and curl) induces a bias of $N^{(1)}$ type (Kesden et al. 2002) on the lensing spectrum of the other component,[25] but this effect is smaller than their individual $N^{(0)}$ noise variance levels.

For the test conducted in the following, we add a Gaussian realization of the post-Born lensing curl component to a simulation. This Gaussian realization comes from Equation (42) —induced power spectrum calculated by CAMB,[26] and is shown as the black line in Figure 5. The lensing curl is second order in the scalar perturbations. A consistent framework would require the inclusion of the corresponding post-Born corrections to the lensing gradient potential (this is, however, smaller than 0.2% and we can therefore neglect it here), as well as the usage of non-Gaussian simulated lensing potentials. However, the latter goes beyond the scope of this paper and we stick to Gaussian realizations, which still provide an important test of the joint reconstruction.

We jointly reconstruct the convergence $\kappa$ and field rotation $\omega$ on the full sky by starting from their Wiener-filtered quadratic estimate.[27] The reconstruction is done for $1 \leqslant L \leqslant 5120$ for $\kappa$ and $2 \leqslant L \leqslant 5120$ for $\omega$. The same lensing curl spectrum used for the simulation generation is used as the prior on the curl modes for the reconstruction in Equation (11). We pick a homogeneous polarization noise level of $0.5\,\mu$K arcmin, use modes of the data up to $\ell_{\max} = 4096$, and reconstruct the unlensed CMB $E$ mode up to $\ell_{\max,\text{unl}} = 5120$.

Figure 5 shows in the upper panel the spectrum of the reconstructed curl in orange, with blue being QE for comparison. We have rescaled the MAP spectrum by the

---

[23] We ignore here nonscalar contributions, such as lensing by gravitational waves, which are expected to result in negligible signals (Cooray et al. 2005).

[24] Contrary claims are given, for example in Cooray et al. (2005) and Pratten & Lewis (2016). However, this is likely due to small numerical inaccuracies in Cooray et al. (2005).
[25] See for example Appendix A of Planck Collaboration et al. (2020a) for the general form of $N^{(1)}$-type contributions.
[26] https://camb.readthedocs.io/en/latest/postborn.html
[27] We use the same Planck lens package to build the curl-mode quadratic estimators.

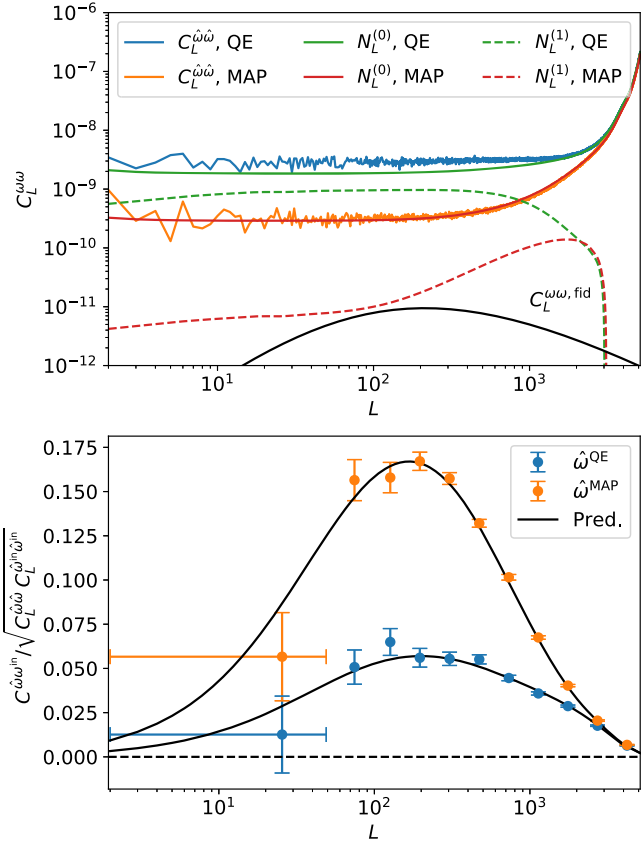

**Figure 5.** Lensing field rotation reconstruction results from our full-sky joint lensing gradient and curl reconstruction. Upper panel: spectra of the curl quadratic estimator and iterative reconstruction are shown in blue and orange, after rescaling by the predicted Wiener filter amplitude, as described in the text. The predicted $N^{(0)}$ biases are shown in green and red. The black line shows the field rotation power spectrum in our fiducial cosmology. The dashed lines show the corresponding $\kappa$-induced $N^{(1)}$ lensing biases (the $\omega$-induced $N^{(1)}$ are much smaller and negligible), which is mostly relevant in the QE case. Lower panel: cross-correlation coefficient of this reconstruction to the input curl modes, together with the predictions built from the $N^{(0)}$ and $N^{(1)}$ biases. Error bars are obtained from the scatter within bins assuming independent estimates at each and every $L$. The auto-spectrum detection S/N jumps by a factor of 5 with the iterative solution.

isotropic normalization, $(\mathcal{W}_L^\omega)^2$, with

$$\mathcal{W}_L^\omega = \frac{C_L^{\omega\omega,\mathrm{fid}}}{C_L^{\omega\omega,\mathrm{fid}} + N_L^{(0),\omega\mathrm{MAP}}}. \tag{43}$$

Regarding the lensing gradient, Legrand & Carron (2022) demonstrate that the approximation of this normalization, and applied to the MAP solution, is accurate within a few percent. This appears to also be a good fit for the curl-mode normalization, and the resulting spectrum can nicely be described by the expected reconstruction noise $N_L^{(0),\omega\mathrm{MAP}}$, as shown in red. In this joint reconstruction, there are two types of $N^{(1)}$ bias: the standard one induced by $\kappa$, and the one induced by the nonzero $\omega$. The latter, however, is much smaller than the $\kappa$-induced bias and is negligible. For the QE, the $N^{(1)}$ bias (green-dashed line) plays a relevant role in the auto-spectrum, but is strongly suppressed for the MAP solution (red-dashed line), since both the lensing $B$ power and lensing power are reduced by large factors. In the lower panel of Figure 5, the points show binned estimates of the empirical cross-correlation coefficient to the input lensing curl modes,

$$\rho_L^\omega \equiv \frac{C_L^{\hat{\omega}\omega^{\mathrm{in}}}}{\sqrt{C_L^{\hat{\omega}\hat{\omega}} C_L^{\omega^{\mathrm{in}}\omega^{\mathrm{in}}}}}. \tag{44}$$

The error bars were calculated from the empirical scatter within each bin, assuming statistically independent fluctuations across multipoles. The black lines show the predictions

$$\rho_L^{\omega,\mathrm{pred}} = \left( \frac{C_L^{\omega\omega,\mathrm{fid}}}{C_L^{\omega\omega,\mathrm{fid}} + N_L^{(0),\omega} + N_L^{(1),\omega}} \right)^{1/2}, \tag{45}$$

for the QE and MAP. The improvement brought by the iterative method is clearly substantial: the full-sky QE S/N,

$$(S/N)^2 \equiv \frac{1}{2}\sum_L (2L+1)(\rho_L^{\omega,\mathrm{pred}})^2, \tag{46}$$

on the auto-spectrum detection is S/N ∼ 2.1 for this configuration, but reaches S/N ∼ 10.4 for the MAP method. Of course, at the polarization noise level of 0.5 $\mu$K arcmin, CMB-S4 would probably observe only a small fraction of the sky, so a purely internal detection in this configuration still remains challenging.

Finally, the impact on the residual lensing $B$-mode amplitude is small. Comparing delensing efficiencies obtained on this same simulation with and without curl modes, we find that the amplitude is reduced by 0.4% compared to the former case. This allows us to conclude that our reconstruction tools perform as expected, even in the presence of the lensing curl. For simplicity, the curl mode is set to zero for the rest of the paper.

### 2.10. A Faster Scheme under Idealized Conditions

The construction of the Wiener-filtered delensed $E$-mode map, as given by Equation (19), is generically the most expensive step of the entire lensing reconstruction procedure. This equation may be rewritten as

$$E_\alpha^{\mathrm{WF}} = C^{EE,\mathrm{unl}}[C^{EE,\mathrm{unl}} + N_\alpha]^{-1} E_\alpha^{\mathrm{delens}}, \tag{47}$$

where $E_\alpha^{\mathrm{delens}}$ is the delensed CMB,

$$E_\alpha^{\mathrm{delens}} \equiv N_\alpha\, _2\mathcal{Y}^\dagger \mathcal{D}_\alpha^\dagger \mathcal{B}^\dagger N^{-1} X^{\mathrm{dat}}, \tag{48}$$

and the prefactor to it is the Wiener-filtering matrix. The interpretation of $E_\alpha^{\mathrm{delens}}$ as the delensed $E$ mode can be justified by noting that the signal part of $E_\alpha^{\mathrm{delens}}$ is the unlensed CMB, provided that the true transfer function matches that of the likelihood model and that $\alpha$ is the true lensing. While still tractable, these optimized weights and filters are not trivial to apply, even under idealized conditions, because $\alpha$ breaks the isotropy of the delensed noise, making $N_\alpha$ a dense matrix. However, this is a small effect, and, under idealized conditions, one can still expect to recover the bulk of the signal with suitable approximations using much faster filtering (as suggested by the pioneering papers Hirata & Seljak 2003a, 2003b). At each iteration, it is quite natural to replace Equation (47) with

$$[E_\alpha^{\mathrm{WF}}]_{\ell m} \to \frac{C_\ell^{EE,\mathrm{unl}}}{C_\ell^{EE,\mathrm{unl}} + N_\ell^{EE}} [\,_2\mathcal{Y}^{-1} \mathcal{D}_\alpha^{-1} \mathcal{B}^{-1} X^{\mathrm{dat}}]_{\ell m}, \tag{49}$$

where the inverse beam and transfer operation, $\mathcal{B}^{-1}$, is a simple rescaling by the isotropic beam, $1/b_\ell$, in harmonic space. $N_\ell^{EE}$ is the isotropic instrument noise, and $_2\mathcal{Y}^{-1}$ is the projection onto the $E$ mode. There is no conjugate-gradient inversion to perform here, which makes this version significantly faster, numerically.

We tested this approximation in the same configuration as in the previous section. At each step, the quadratic gradient calculation takes the slightly suboptimal Wiener-filtered $E$ mode as input, but is otherwise unchanged. We found that the first iteration worsened the solution significantly below $L = 30$, which we solved simply by fixing these large lenses to the QE solution. On the other hand, smaller scales appear to be recovered almost equally well. After 15 iterations, we found that the cross-correlation coefficient is within 1% of the solution without these approximations for $100 \leqslant L \leqslant 5120$. While we do not pursue this further here, it clearly means that the simplified gradient calculation can be valuable for theoretical investigations using idealized conditions. Work is ongoing to understand to what extent faster yet efficient methods can also be devised in realistic configurations.

## 3. Simulations

This section discusses the individual components of the simulated maps and how they are generated and combined.

### 3.1. CMB Signal Simulations

The simulations used in this work are generated using the lensed CMB maps from the Planck FFP10 simulation suite as input. These CMB maps, in harmonic space, were built by one of the corresponding authors in preparation for the 2018 Planck release, and were later made public on NERSC and used by the CMB-S4 Collaboration. The maps contain a series of small lensing-related defects (and another one was also introduced at the very beginning of this work), which are completely innocuous at most noise levels and on large scales relevant for $r$ inference. However, in precise tests performed at low polarization noise levels, these defects occasionally caused troubles. In the following, we provide a record of these issues along with the relevant lensing properties.

1. For the simulations, the rotation of the polarization that is required to match the change in reference axes caused by the sky curvature (the phase in Equation (2)) was not performed. As a consequence, the polarization right at the poles is just wrong since this phase is not small. This is something to which our reconstruction pipeline reacts (for noise below the $\mu$K arcmin level) by assigning absurd values to the lensing potentials at the poles, or failing to converge altogether. However, only a few arcmins away from the poles, we find that this is not an issue any longer.
2. Due to a bug in the interpolating code that was used at the time, a set of about ∼20 pixels at $N_{\text{side}} = 2048$ that are located in the southern hemisphere, well separated from one another, and close to the meridian defined by $\varphi = 0°$, were just wrong.
3. In addition to the massive CMB dipole, our velocity relative to the CMB frame also causes more subtle modulation and aberration of the anisotropies along the velocity direction (Planck Collaboration et al. 2014). This aberration results in an almost perfect dipole lensing signal, approximately five times larger than the expected $\Lambda$CDM lensing potential dipole, and is easily detected with reasonably wide sky coverage (Planck Collaboration et al. 2014). Utilized for this were CMB frequency maps that contain the frequency-independent part of the modulation plus the aberration, for all Planck channels.
4. After synthesizing the FFP10 lensed $E_{\ell m}$ and $B_{\ell m}$ into real space maps, the CMB-S4 team adds foreground maps defined in equatorial coordinates to these maps. However, the Planck team worked in Galactic coordinates, and this leads to the aberration signal mentioned just above having an inconsistent direction to the pair of statistically anisotropic galactic foreground models that we will be using.

### 3.2. Instrument Noise Simulations

Noise is generated in the same manner as was done in our previous CMB-S4 $r$-forecast paper (Abazajian et al. 2022). Uniform full sky realizations of noise are generated and then divided by the *Pole Deep* and *Chile Full* hits patterns, which are shown in Figure 7 of that paper. The normalization of the noise is adjusted such that when reanalyzing these realizations with inverse-variance weighting one obtains the power spectrum noise levels given in Table 1. These hit patterns come from full-blown time-ordered-data level scanning simulations for the CMB-S4 small aperture telescopes (SATs). The noise levels in Table 1 come from a preliminary set of CMB-S4 measurement requirements. For the present we assume that the large aperture telescopes (LATs) have the same hits pattern—as could be (approximately) arranged in practice. Excess low-frequency $1/f$ noise is included with the knee and slope parameters given in the table. These parameters are derived from BICEP/Keck and South Pole Telescope (SPT) data (Abazajian et al. 2022). Map noise contours are shown in Sections 4.2.1 and 4.3. Experiments at the South Pole, or more generally ground-based experiments, can suffer from large-scale noise that is highly anisotropic in harmonic space, sourced by atmospheric noise and the scanning strategy. This type of anisotropy is not included in this series of simulations and the study of its impact on the iterative lensing map is left for future work.

### 3.3. Foreground Models

We run the curved sky lensing reconstruction algorithm on three different simulation sets with varying degrees of foreground complexity. We refer to the three sets containing foreground and CMB only as *Sky Model 00, 07, 09*, and as the *Simulation Set* ($M_{00}$, $M_{07}$, $M_{09}$), if it includes all components (foregrounds, CMB, and noise). In our previous paper (Abazajian et al. 2022), more foreground models were studied. However, the other models did not extend to the small scales relevant to the lensing reconstruction.

$M_{00}$ includes the simplest possible (and completely unrealistic) foreground model; Gaussian realizations of synchrotron and dust with uniform amplitude across the full sky. For the synchrotron $A_s$, and dust amplitude $A_d$, we use $A_s = 3.8 \, \mu\text{K}^2$ and $A_d = 4.25 \, \mu\text{K}^2$, for the corresponding $\alpha$ parameters, we use $\alpha_s = -0.6$ and $\alpha_d = -0.4$, and for the $\beta$ parameters $\beta_s = -3.1$ and $\beta_d = 1.6$. For the temperature of the dust $T_d$, we choose $T_d = 19.6$ K (see BICEP2 Collaboration et al. 2018 for a description of the foreground emission laws). There are good reasons to include such a model in this study; the mean field from this simulation set will not pick up foreground residuals (other than foreground inhomogeneities arising from the magnifying and demagnifying of the delensing procedure, analogous to the induced noise inhomogeneity, as discussed in Section 2.3), which will be useful for assessing the impact of the mean field on our results.

Table 1
Simulated SAT and LAT Map Noise Levels

| | Band (GHz) | Beam (arcmin) | EE White ($\mu$K arcmin) | EE $\ell$-knee | EE Slope | BB White ($\mu$K arcmin) | BB $\ell$-knee | BB Slope |
|---|---|---|---|---|---|---|---|---|
| SAT | 30 | 72.8 | 3.74 | 60 | −2.2 | 3.53 | 60 | −1.7 |
| | 40 | 72.8 | 4.73 | 60 | −2.2 | 4.46 | 60 | −1.7 |
| | 85 | 25.5 | 0.93 | 60 | −2.2 | 0.88 | 60 | −1.7 |
| | 95 | 22.7 | 0.82 | 60 | −2.2 | 0.78 | 60 | −1.7 |
| | 145 | 25.5 | 1.25 | 65 | −3.1 | 1.23 | 60 | −3.0 |
| | 155 | 22.7 | 1.34 | 65 | −3.1 | 1.34 | 60 | −3.0 |
| | 220 | 13.0 | 3.48 | 65 | −3.1 | 3.48 | 60 | −3.0 |
| | 270 | 13.0 | 8.08 | 65 | −3.1 | 5.97 | 60 | −3.0 |
| LAT | 20 | 11.0 | 13.16 | 150 | −2.7 | 13.16 | 150 | −2.7 |
| | 30 | 7.3 | 6.50 | 150 | −2.7 | 6.50 | 150 | −2.7 |
| | 40 | 5.5 | 4.15 | 150 | −2.7 | 4.15 | 150 | −2.7 |
| | 95 | 2.3 | 0.63 | 150 | −2.6 | 0.63 | 150 | −2.6 |
| | 145 | 1.5 | 0.59 | 200 | −2.2 | 0.59 | 200 | −2.2 |
| | 220 | 1.0 | 1.83 | 200 | −2.2 | 1.83 | 200 | −2.2 |
| | 270 | 0.8 | 4.34 | 200 | −2.2 | 4.34 | 200 | −2.2 |

$M_{07}$ includes an amplitude-modulated foreground model; we take the Gaussian fields from $M_{00}$ and modulate their amplitude across the sky according to a map of degree-scale $B$-mode power measured from small patches of Planck's 353 GHz map in a similar manner to Figure 8 of Planck Collaboration et al. (2016).

$M_{09}$ includes a realization of the multilayer Vansyngel model (Vansyngel et al. 2017). Each layer has the same intensity (constrained by the Planck intensity map), but different magnetic field realizations. It produces $Q$ and $U$ maps by integrating along the line of sight over these multiple layers of magnetic fields.

The complete simulation sets are generated as follows. For each, we sum the noise, lensed CMB, and the respective foreground for each simulated frequency band. All components are generated in equatorial coordinates, with the exception of the FFP10 lensed CMB. This is not expected to pose an issue for lensing reconstruction, as discussed in the previous section. The component $fg_{09}$ is rotated to equatorial coordinates before adding it. Each realization contains a different realization of the lensed CMB and noise. For $M_{00}$ and $M_{07}$, each realization contains a different foreground realization, with the amplitude modulation of $fg_{07}$ being the same in each realization. For $M_{09}$, there is only one realization; each and every map therefore contains the exact same foreground component. The foreground models do not contain extragalactic foregrounds, and our analysis focuses on polarization data only. Extragalactic polarized foreground contamination is generally expected to be weak and not to affect the lensing reconstruction significantly, dominated by the $EB$ signal. Source masking can potentially bias the lensing reconstruction if the mask cross correlates substantially to the lensing signal. However, this has also been shown to be small (?) in our configuration. Therefore, we do not include point-source masks in our analysis.

## 4. Results

The discussion on the full pipeline of the $r$-analysis and the analysis of the robustness of our iterative curved sky lensing reconstruction is split as follows: in Section 4.1, we discuss the component separation technique used to produce a clean CMB map from the LAT simulated maps. In Sections 4.2 and 4.3, we discuss the performance of the lensing reconstruction and the generation of the $B$-lensing template to obtain the best estimate of the $B$-lensing signal. In Section 4.4, we discuss the construction of the likelihood for $r$ and resulting constraints.

### 4.1. Component Separation

We apply a harmonic internal linear combination (ILC) on the LAT frequency maps (Tegmark et al. 2003). We use this for all frequencies with the exception of the 20 GHz channel, which is placed on the LAT but has the primary function of aiding in the constraint of the synchrotron component in the likelihood for $r$; the beam of the 20 GHz channel is too large to make any meaningful contribution for the purpose of lensing reconstruction and is therefore not taken into account for the LAT component separation. We obtain the weights $\omega_\ell$ analytically by creating a covariance matrix $R_\ell$ and applying the formula,

$$\omega_\ell = \frac{a^T R_\ell^{-1}}{a^T R_\ell^{-1} a}, \quad (50)$$

where $a$ is the emission law of the CMB in the chosen units. The covariance matrix $R_\ell$ is composed of input theoretical spectra and cross spectra of simulation set $M_{00}$ maps. We do this for the $E$ mode and $B$ mode independently, but apply a single set of weights to all simulation sets. These weights are derived analytically from the input power spectra of the CMB + noise + $fg_{00}$. We compare the weights obtained from the input spectra, and from the spectra obtained from the realizations and find that for each simulation set and on the scales relevant to the lensing reconstruction they differ by a margin we consider acceptable. At most, the weights between the input and empirical spectra varied 10%. The above approach is, by no means, an optimized approach, and more sophisticated component separation techniques are currently studied. The weights are applied as follows. We first convert the Stokes parameters $Q$ and $U$ to $E$ and $B$ with a standard spin-2 transform using the Healpix package (Górski et al. 2005), and rebeam the maps to a common resolution of $2'\!.3$ FWHM. In order to avoid too large $E$-to-$B$ leakage, we first multiply the maps with an apodization mask. Here we simply use the hits map of the scanning strategy, smoothed by a $0°\!.5$ Gaussian beam. We then isotropically weight the harmonic coefficients and reproject them onto $Q$ and $U$. Owing to masking, the resulting maps are suppressed close to the patch edges. The lensing estimation pipeline always assumes that the local response of the data map to the CMB is isotropic. We approximately account for this by simply rescaling the ILC maps by the inverse of the mask.

An estimate of each of the foreground model residuals is shown in Figure 6, and is calculated by combining all frequency maps with the ILC weights from the component separation step, and for the foreground-only maps. We also indicate the boundaries of the South Pole Deep and Chile Full hits pattern indicated by the dark red and gray contours.

Figure 7 shows the $EE$ ($BB$) noise, CMB, and foreground power spectra estimates on the South Pole Deep Patch (SPDP) in the top (bottom) panel and for a single realization. The fiducial CMB signal is shown in red. The noise is shown in blue, and becomes the dominant component at $L \gtrsim 4000$ in $E$, and at $L \gtrsim 2800$ in $B$. We also show a theoretical purely white-noise power spectrum of $0.45\,\mu$K arcmin in thin gray. The different foregrounds fg$_{00}$ (dashed–dotted line), fg$_{07}$ (solid), and fg$_{09}$ (dashed line) are shown in green; The component fg$_{07}$ has the largest amplitude in both $E$ and $B$ and on most scales. The black data points show the binned power spectra of the component-separated maps, and the error bars indicate the uncertainties on the bins.

### 4.2. SPDP Map-based Delensing

We now discuss the QE and iterative lensing reconstruction, which we use to delense the simulations on the SPDP. We use 500 ILC component-separated maps of simulation sets M$_{00}$, M$_{07}$, and M$_{09}$ on the SPDP, which covers about 5% of the sky. Our final products are QE and MAP method Wiener-filtered $E$-maps, QE and MAP method lensing potential maps, and QE and MAP $B$-lensing templates. We quantify the performance and accuracy of our $B$-lensing templates by calculating the residual lensing amplitude $A_{\rm lens}$ and split the SPDP into various sky patches of different noise levels. We analyze $A_{\rm lens}$ as a function of the sky patches that we consider, and analyze the robustness of the lensing reconstruction by changing parts of the pipeline and dependencies. If not stated otherwise, the calculation of a power spectrum is always done with our own implementation of the PolSpice (Chon et al. 2004) algorithm.

#### 4.2.1. B-lensing Templates

As discussed in detail in Section 2.2, the reconstructions (both the QE and iterative solution) take a number of fiducial ingredients as input, notably a suitable noise covariance matrix model. We do this by calculating a central noise level $\hat{n}_{\rm lev}$ and scale it with the hits count map as follows. To obtain $\hat{n}_{\rm lev}$, we calculate the empirical power spectrum of the component-separated noise maps for the central area of the patch, $\hat{C}_\ell^{\rm noise}$, and

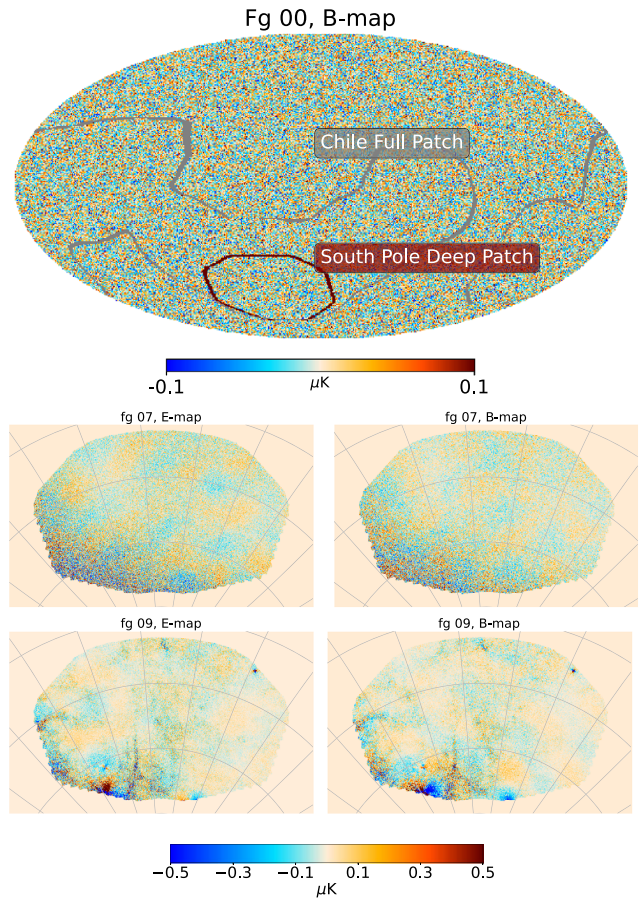

**Figure 6.** Foreground residual maps for fg$_{00}$ (top), fg$_{07}$ (center), and fg$_{09}$ (bottom), bandpassed with $10 < \ell < 3000$, in equatorial coordinates obtained via combining the individual foreground-only frequency maps with the ILC weights. For fg$_{00}$, we only show the $B$ map in the Mollweide projection, whereas for fg$_{07}$ and fg$_{09}$ we show the $E$ map (left column), and the $B$ map (right column) in the Cartesian projection and about the SPDP. The graticules in the Cartesian projection plots have a spacing of 15° in both the R.A. and decl. directions. The edges of the SPDP are indicated by the red honeycomb shape in the top panel, and the gray wiggly line indicates the Chile Full Patch.

then convert that result to a white-noise level in $\mu$K arcmin, using

$$\hat{n}_{\rm lev} = \sqrt{\langle \hat{C}_\ell^{\rm noise}\rangle_\ell}\,\frac{60\cdot 180}{\pi}, \quad (51)$$

where $\langle\rangle_\ell$ denotes the average over the multipole range $1800 < \ell < 2000$. We choose this range (which is not particularly critical) because lensing reconstruction utilizes information on the CMB at the smallest scales possible, which are, at the same time, not too noisy. While the $E$ mode is always close to being effectively noise-free, this range is roughly at the scales at which, for the first iterations in the $B$-mode map, noise becomes the dominant component (see Figure 7). With this, we find $\hat{n}_{\rm lev} = 0.42\,\mu$K arcmin. We scale this number with the inverse root of the hits count map to obtain our fiducial inhomogeneous noise map across the full patch.[28] The resulting SPDP noise variance map is shown in Figure 8, which ultimately is used for both reconstruction methods. The procedure described above does not account for the residual foregrounds in the maps, and therefore, results in slightly suboptimal reconstructions. However, the analytical forecast tool described earlier predicts this to be a small effect[29] and we neglect it in this paper. We also neglect any effective noise correlation introduced by the component separation procedure, foregrounds, or $1/f$ noise.

---
[28] For a real data analysis, the same simulation-based approach might not necessarily be as accurate, but the precise value is not critical and other ways are possible, for example, using splits of the data.

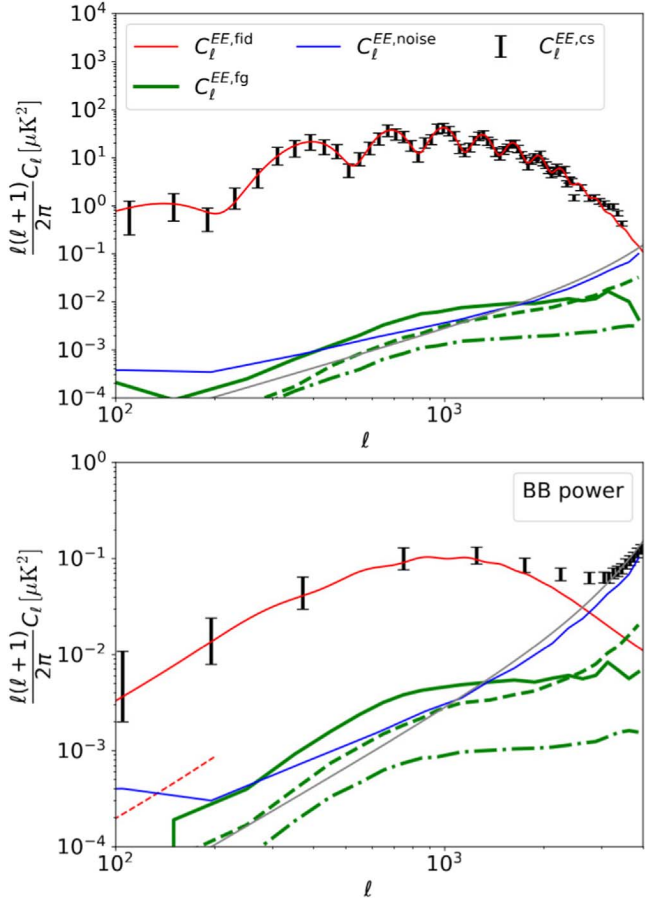

**Figure 7.** Binned LAT residual power spectra on the SPDP for a sky fraction of about 2% about the central area. The component-separated $M_{07}$ CMB (black) with a simple error estimation shown as error bars, foreground residuals (green) for foreground model $fg_{00}$ (dashed–dotted line), $fg_{07}$ (solid), and $fg_{09}$ (dashed line), and noise (blue line) are shown in both panels. The component $fg_{07}$ has the largest power compared to all foreground models. The gray line shows the theoretical white-noise-level power spectrum of 0.45 $\mu$K arcmin and is the estimated average of that patch (see Figure 8). Top panel: $E$-mode power spectra. The red line represents the fiducial lensed $E$-mode power spectrum. Noise becomes the dominant component at about $\ell = 4000$. Bottom panel: $B$-mode power spectra. The red line represents the fiducial $B$-lensing power spectrum, and the red-dashed line shows the theoretical $B$ power spectrum with a residual lensing amplitude of 6% (we only show multipoles $\ell < 200$, as these are the ones that our procedure can produce). Noise becomes the dominant component at about $\ell = 2800$.

Our fiducial model transfer function is the same isotropic Gaussian beam with an FWHM of $2\rlap{.}'3$ used for the component separation, along with the appropriate pixel window function, and the fiducial spectra for $C_L^{\kappa\kappa,\mathrm{fid}}$ and $C_\ell^{EE,\mathrm{unl}}$ are the input FFP10 cosmology spectra.

The recovery of the unlensed $E$ modes with $\ell_E < 10$ is not attempted in the baseline reconstructions. This is taken care of by setting the fiducial transfer function in the likelihood for these multipoles to zero. We apply the overlapping $B$-mode deprojection on the noise model as discussed in Section 2.7 for scales $\ell_B \leqslant 200$, suppressing all these $B$ modes by a factor of $10^4$. Further, a prior much smaller than $C_L^{\kappa\kappa,\mathrm{fid}}$ is set on the lenses with $L < 4$. This is done to reduce the size of the Newton steps for these scales and for all iterations, so that the algorithm becomes numerically more stable on these scales.

---

[29] In a comparable experimental configuration, we found a predicted relative increase in residual lensing $B$-mode power of 3% to 12% depending on the foreground model.

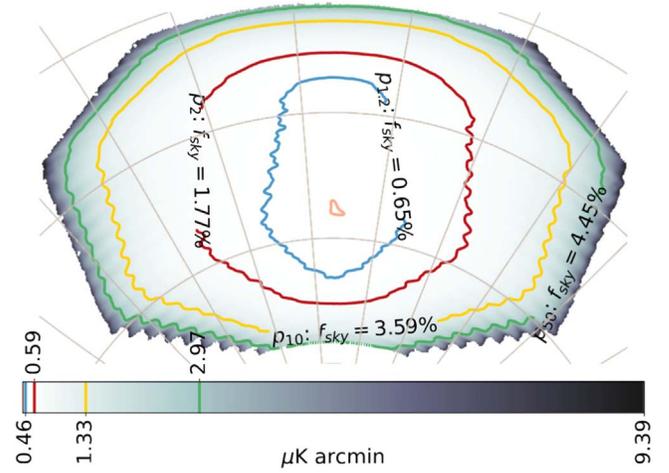

**Figure 8.** SPDP noise-level map used in the lensing likelihood model, derived from calculating a central noise level and scaling it according to the hits count map. The contours show edges that contain the masks $p_{1.2}$ (blue), $p_2$ (red), $p_{10}$ (green), and $p_{50}$ (yellow). Each mask covers the full area inside of it, for example, $p_{1.2}$ is contained in $p_2$. The lowest (highest) noise of about 0.42 (9.39) $\mu$K arcmin is indicated by the pink (dark gray) contour. The edge pixel noise levels of the contours are indicated by lines in the color bar; the pink contour is not shown. The graticules have a spacing of 15° in both the R.A. and decl. directions.

The quadratic estimator is built in the same way as for the latest Planck analyses (Planck Collaboration et al. 2020b; Carron et al. 2022), using Plancklens. The QE mean field $\hat{\alpha}^{\mathrm{QE,MF}}$ is calculated by averaging over 200 reconstructions of $\hat{\alpha}^{\mathrm{QE}}$, less than the available 500 simulations for historical reasons.[30] The QE lensing potential is mean field subtracted, and then isotropically[31] weighted by

$$\mathcal{W}_L^\kappa = \frac{C_L^{\kappa\kappa,\mathrm{fid}}}{C_L^{\kappa\kappa,\mathrm{fid}} + N_L^{(0),\kappa}}, \qquad (52)$$

and this serves as the starting point for the iterative method. The QE uses CMB multipoles up to 3000, but the MAP iterations use modes up to 4000.[32] We assume that the unlensed $E$ mode is band limited to 4000, and neglect the transfer of power caused by lensing beyond this scale.

The mean field part of the gradient, $g_\alpha^{\mathrm{MF}}$ is set to $\hat{\alpha}^{\mathrm{QE,MF}}$ at each iteration, and therefore, kept invariant across iterations.

---

[30] This is not critical; results using 100 simulations only for the QE mean field are extremely close to our baseline of 200.
[31] We do not include the QE a posteriori inhomogeneous weighting called $\kappa$ filtering (Mirmelstein et al. 2019), which could in principle slightly improve the reconstruction and QE lensing templates close to the mask edges.
[32] The values for QE and MAP are different only for historical reasons. The main purpose of the QE estimate is to have a starting point for the MAP estimator; using more multipoles to build the QE keeps the converged MAP solution unchanged. In the baseline configuration discussed in this paper, the predicted residual lensing amplitude for QE delensing reduces from 23.2 to 20.6 if the multipole range is increased to 4000, which could have been done without any caveats.

By doing this, we neglect the delensed noise mean field as discussed earlier, as well as the changes in the mask and noise mean fields induced by the changes in the power of the maps through the iterations. Later in Section 4.2.4, we show that this is an adequate approximation.

During each iteration, we must perform a number of deflection operations $\mathcal{D}_\alpha$. This remapping, in principle, is done on the full sky. However, the reconstructed lensing map is very weak outside of the SPDP. It is therefore a valid simplification to neglect the deflection field sufficiently far away from the SPDP. We define this region to be larger by 5° than the SPDP itself. Outside that region, we effectively assume that the Wiener-filtered, unlensed $E$ mode is zero. We perform the lensing remapping on a reduced Gauss–Legendre geometry (Reinecke & Seljebotn 2013) using a bicubic spline interpolation on an equidistant, cylindrical grid of resolution $1'.7$. This resolution is not always sufficient for precise recovery of the largest lenses, but is good enough for our purposes. A higher resolution increases the robustness of the reconstruction at these low multipoles at the expense of an increased computational cost.[33]

We calculate the $B$-lensing templates following Equation (33), and for the quadratic estimator with the perturbative version of Equation (34), given by Equation (6). Figure 9 shows an example $B$-lensing map before (top panel), and the residual $B$-lensing maps after QE (middle panel), and MAP delensing (bottom panel) on the patch area. As expected, most power is removed at the center of the patch where the data noise is lowest and therefore lensing reconstruction performs best.

### 4.2.2. Delensing Performance

We now characterize the fidelity of the obtained $B$-lensing templates and quantify the reduction of the $B$-lensing signal after internal delensing in terms of the residual lensing amplitude $A_{\rm lens}$. The QE and MAP residual $B$-lensing maps are calculated by subtracting the respective $B$-lensing templates from the $B$-lensing maps by using Equation (33). We obtain $A_{\rm lens}$ by calculating the ratio between the $B$-lensing power spectrum and the residual $B$-lensing power spectrum, and compare our results to the analytical predictions obtained by our tools described earlier in Section 2.8.

Both the noise and foreground residuals are inhomogeneous across the patch. For this reason, the quality of the lensing reconstruction varies across the observed area. To quantify the residual lensing amplitudes as a function of position, we define sky patches inside the SPDP following its hits count map: patch $p_t$ with hits count ratio threshold $t$ includes the observed pixel $i$ with the hits count $h_i$, if

$$h_i < \frac{1}{t}. \quad (53)$$

Here, $h_i$ is normalized by the maximum number of hits. The conversion between $h_i$ and its noise level $n_{{\rm lev},i}$ is

---

[33] As we neared the completion of this paper, our lensing remapping procedure has seen major improvements (Reinecke et al. 2023) in accuracy and execution time, allowing better recovery of the large-scale lenses and removing the need for the simplifications just discussed. While the entire set of maps of this analysis has not been reprocessed, dedicated tests on M$_{00}$ are indicating that our latest and most ambitious version of the code can lead to visible improvements, reaching a residual lensing amplitude of 0.059, compared to 0.069 obtained in this paper in Section 4.2.2.

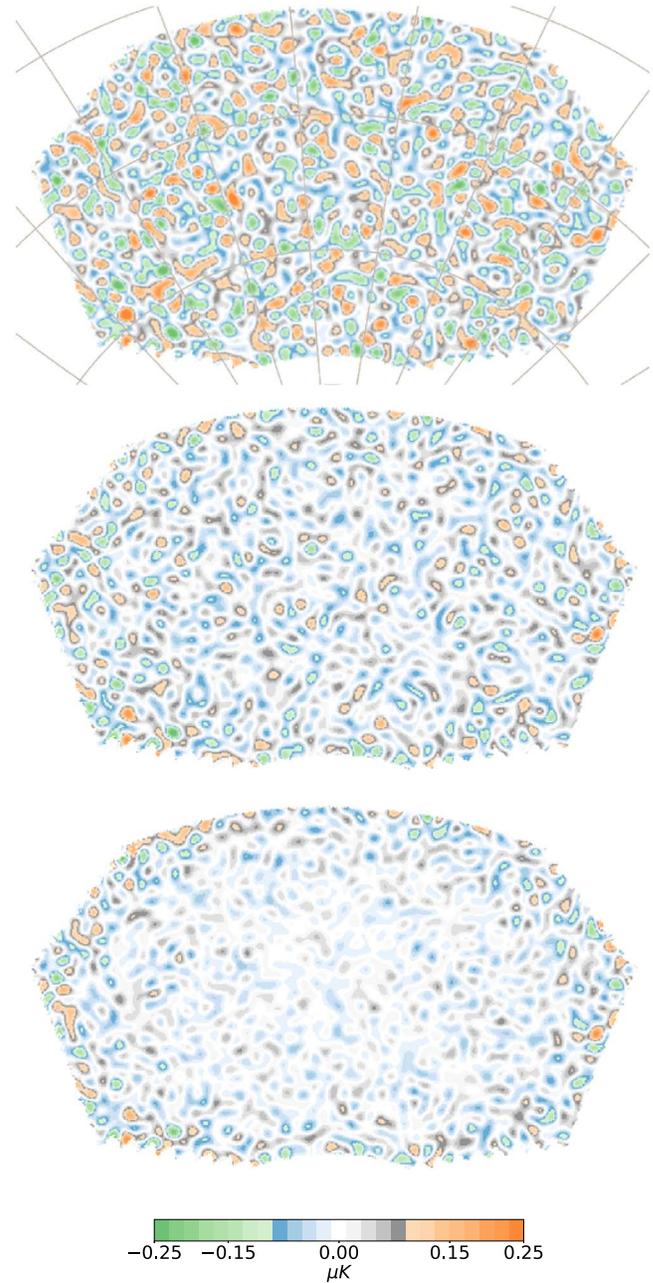

**Figure 9.** CMB $B$-lensing maps of the SPDP, before (top panel) and after quadratic estimator (middle panel) and iterative delensing (bottom panel), using the $B$-lensing templates from simulation set M$_{00}$. The maps are bandpassed to only show $30 < \ell < 200$. Delensing works best in the center of the patch, where the improved performance of the iterative solution is clearly visible. Close to the mask edges the instrumental noise increases, reducing the quality of the lensing reconstruction and of the delensing. The graticules have a spacing of 15° in both the R.A. and decl. directions.

straightforward, if the central noise level $\hat{n}_{\rm lev}$ is known:

$$n_{{\rm lev},i} = \frac{\hat{n}_{\rm lev}}{\sqrt{h_i}}. \quad (54)$$

Thus, an area of the sky observed 10 times less frequently is about 3.3 times noisier. We choose $t = [1.2, 2, 10, 50]$ and obtain patches with sky fractions between 0.07% and 4.4% (see Figure 8). The blue patch is the essentially homogeneous

**Table 2**
Average and Maximum Noise Levels for the Patches Chosen in This Analysis

| Identifier | Contour Color | $f_{\rm sky}$ (%) | Average Noise Level ($\mu$K arcmin) | Maximum Noise Level ($\mu$K arcmin) |
|---|---|---|---|---|
| $p_1$ | Pink | … | 0.42 | 0.43 |
| $p_{1.2}$ | Blue | 0.65 | 0.44 | 0.46 |
| $p_{2.0}$ | Red | 1.77 | 0.48 | 0.59 |
| $p_{10}$ | Yellow | 3.59 | 0.55 | 1.33 |
| $p_{50}$ | Green | 4.45 | 0.64 | 2.97 |
| $p_\infty$ | … | 4.97 | 0.67 | 9.39 |

**Note.** $f_{\rm sky}$ is the ratio of the number of considered pixels to the total in the patch. See the text for a discussion.

central region, with an average noise level of 0.44 $\mu$K arcmin, and its edge pixels noise level is about 0.46 $\mu$K arcmin. This is indicated as a blue line inside the color bar at the bottom. A summary of the average and maximum noise levels together with its sky fraction and the hits count ratio thresholds is shown in Table 2. The B-lensing signal is about 5 $\mu$K arcmin. With a central noise level of $\hat{n}_{\rm lev} = 0.42$ $\mu$K arcmin, we therefore expect the iterative lensing analysis to become ineffective at ~10 times the central noise value, as B lensing is an ~5 $\mu$K arcmin contamination that cannot be resolved anymore at these levels. This boundary is close to the green contour that can be seen in Figure 9, where delensing is much weaker and the iterative method brings little improvement.

The power spectra of the B-lensing and residual B-lensing maps of M$_{07}$ for each of the sky patches are shown in Figure 10, using a binning window of $\Delta\ell = 30$ for the calculation. Each panel shows the ensemble average and standard deviation of the B lensing (orange), QE (dark blue), and MAP (light blue) delensed power spectra for the patch indicated by the legend referring to Figure 8. The orange line shows the input B-lensing signal, and the blue lines show the analytic predictions of the delensed power spectra. Later, we provide more details about the calculation of the predictions. As can be seen, the delensed power spectra are largely scale independent and increase with increasing sky fraction due to the increase in noise. At the same time, the standard deviation decreases due to the larger sky fraction reducing the sample variance. We do not show the first bin of the delensed power spectra, since the simulations do not contain simulated signals on the reionization peak.[34]

Across all simulation sets, M$_{07}$ gives the highest ensemble variance, and is consistent with M$_{07}$ having the largest foreground residuals, see Figure 7.

We now calculate $A_{\rm lens}$ as the ratio of the ensemble mean of the power spectra using the relation between the power spectrum and its Gaussian variance. This approximates the decrease in the B power fluctuations, since for Gaussian fields, the spectrum variance $(\Delta C_\ell^{BB})^2$ is quadratic in the power itself,

$$(\Delta C_\ell^{BB})^2 \approx \frac{2(C_\ell^{BB})^2}{f_{\rm sky}(2\ell+1)}, \qquad (55)$$

---
[34] By construction, the CMB-S4 simulation suite used in this work did not contain signals below $\ell = 30$, in order to force conservative forecasts based on the recombination peak signal only.

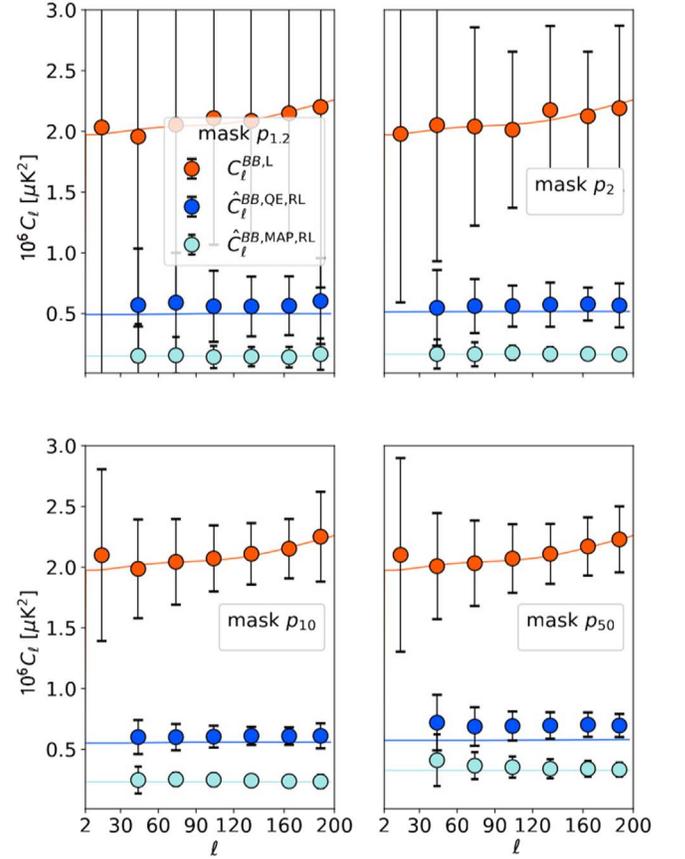

**Figure 10.** Ensemble average and standard deviation of the B lensing (orange), QE delensed (dark blue), and MAP delensed (light blue) B-lensing power spectrum, binned with $\Delta\ell = 30$, for simulation set M$_{07}$. The power spectra are calculated on four different sky patches (see Figure 8), one per panel. The solid lines show the fiducial B-lensing signal (orange), and analytical prediction for the QE (dark blue) and MAP residual power spectrum. Delensing performs almost equally well across all scales but shows a dependence on the sky patch owing to inhomogeneous noise and foreground residuals. Results on the other foreground models are qualitatively similar. See Figure 11 for more detailed comparisons.

where $f_{\rm sky}$ is the sky fraction from which the power spectrum is computed. We quote the QE and MAP delensed $A_{\rm lens}$ using

$$A_{\rm lens}^{\rm QE/MAP} = \frac{\langle C_\ell^{BB,{\rm QE/MAP,RL}}\rangle_{30\leqslant\ell\leqslant 200}}{\langle C_\ell^{BB,{\rm L}}\rangle_{30\leqslant\ell\leqslant 200}}, \qquad (56)$$

where $C_\ell^{BB,{\rm QE/MAP,RL}}$ ($C_\ell^{BB,{\rm L}}$) is the (residual) B-lensing power spectrum. The results are shown in Figure 11. The left (right) panel shows the empirical MAP (QE) result and all three simulation sets in light blue (dark blue), where M$_{00}$ is shown in the top two columns, M$_{07}$ in the middle two columns, and M$_{09}$ in the bottom two columns. The error bars display the standard deviation of this ratio across the simulation set.

Each panel shows the empirical results on $\hat{A}_{\rm lens}$ for each sky patch, and includes semi-analytical predictions (orange and pink), which we discuss in the next section. The QE results show the values obtained via perturbative remapping of the E mode, Equation (6). We note that perturbative lensing remapping results in lower residual lensing amplitudes by about 4% compared to nonperturbative results, for all sky patches and simulation sets, in agreement with Baleato Lizancos et al. (2021a). Table 3 reproduces the numbers for the central, most relevant region.

Unsurprisingly, M$_{00}$ gives the lowest values for both QE and MAP, since it has the lowest foreground contamination, and in addition, contains homogeneous Gaussian noise across the SPDP. $M_{07}$ contains the highest foreground contamination, therefore resulting in the highest $A_{\text{lens}}$. $M_{09}$ is the most complex of all three simulation sets, but has slightly less foreground residual power compared to $M_{07}$, leading to a slightly lower residual lensing amplitude.

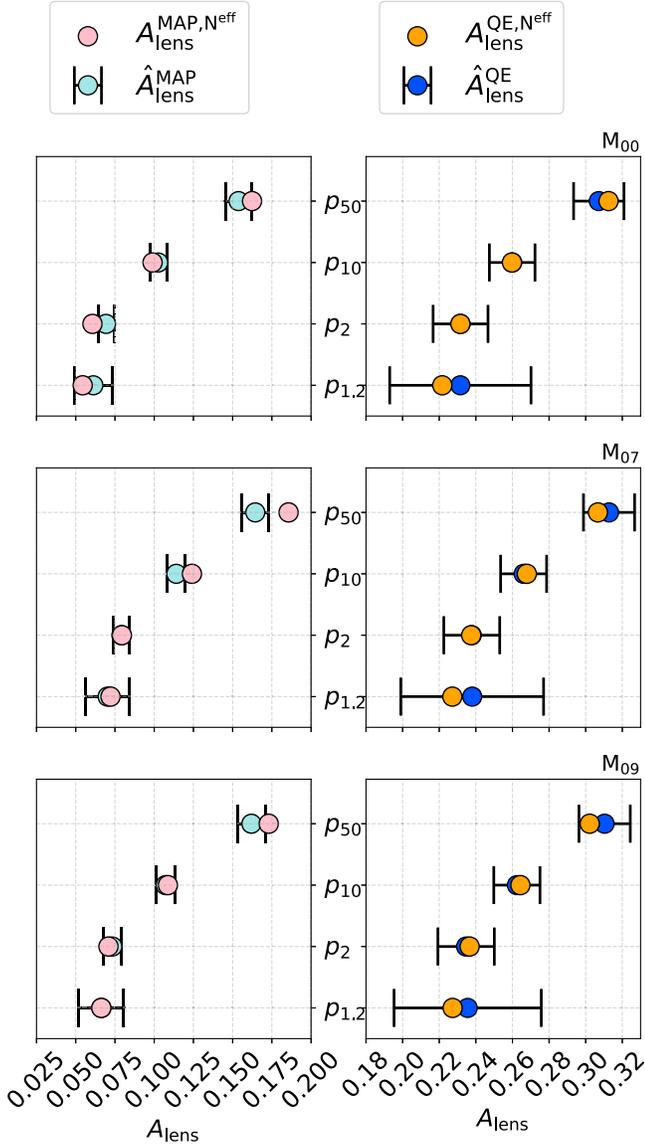

**Figure 11.** Ensemble average and standard deviation of the QE (right panels, dark blue), and MAP (left panels, light blue) residual lensing amplitude $A_{\text{lens}}$ binned across the recombination peak, for $M_{00}$ (top two panels), $M_{07}$ (middle two panels), and $M_{09}$ (bottom two panels), and predicted residual lensing amplitudes shown as pink (MAP) and orange (QE) data points. The dark blue data points are the empirical results using perturbative lensing remapping. In each panel, the data point across the y-axis shows the result for the different sky regions, as indicated by the labels on the y-axis (see Figure 8). $M_{07}$ is always slightly larger than $M_{09}$, and we find the lowest ensemble mean $A_{\text{lens}}$ for MAP delensing in patch $p_{1.2}$. The lowest ensemble standard deviation, however, comes from patch $p_2$. All predictions consider foreground power in the noise estimate.

*4.2.3. Comparison to Predictions*

We now discuss the calculation of our predictions, starting with simple assumptions, and gradually adding components to increase realism.

**Table 3**
Residual Lensing Amplitudes for $30 \leqslant \ell \leqslant 200$ Found in This Study on Region $p_2$ of the SPDP, See Figure 8

|  | $M_{00}$ | $M_{07}$ | $M_{09}$ |
|---|---|---|---|
| $A_{\text{lens}}^{\text{QE}}$ | $0.232 \pm 0.015$ | $0.238 \pm 0.015$ | $0.235 \pm 0.015$ |
| $A_{\text{lens}}^{\text{MAP}}$ | $0.069 \pm 0.005$ | $0.079 \pm 0.005$ | $0.073 \pm 0.005$ |

In the most optimistic case, we can calculate predictions based solely on the central white-noise level of 0.42 $\mu$K arcmin, and ignore foregrounds, sky cuts, and the scan strategy of the experiment. By using the same cuts as in the reconstruction, $(200 < \ell_B < 4000, 10 \leqslant \ell_E \leqslant 4000)$ for MAP, and $(200 < \ell_B < 3000, 10 \leqslant \ell_E \leqslant 3000)$ for QE, we predict $A_{\text{lens}}^{\text{MAP}} = 0.053$ for the polarization-only iterative delensing estimator and $A_{\text{lens}}^{\text{QE}} = 0.23$ for the QE case. For sky model $M_{00}$ on the sky patch $p_2$ we find 0.069 and 0.23, see Table 3. This indicates that in contrast to the QE case, foreground residual power plays a nonnegligible role overall in the iterative reconstruction. This is of course expected to some extent since the size of the foreground residuals relative to the B-lensing power is much greater for the MAP reconstruction.

We can attempt to take foreground residuals into account by assuming they effectively act as additional noise. Foreground models $fg_{07}$ and $fg_{09}$ contain some level of non-Gaussianity, which can in principle affect the reconstruction in less trivial ways. In the baseline reconstructions, this non-Gaussianity is removed by subtracting the mean field from the lensing potential estimates (recall that for these foreground models the same non-Gaussian realization enters all simulations). However, the next section shows that iterative reconstructions, for which the mean field does not match the input foreground model, gives the almost same results.

We construct residual foreground maps by combining the single-frequency simulation-input foreground maps with the weights derived from the component separation step in the previous section, and with the appropriate beams of each detector. By repeating the exercise of calculating the power spectrum for each patch on the sky for an appropriate binning, we obtain the corresponding residual foreground power spectra.

We refer to the input noise in our prediction as *noise only* when we calculate the noise power spectrum from the empirical noise maps, and as *effective noise* if we add the foreground residuals' power spectrum on top of it. Calculating the power spectrum across a sky patch effectively gives the average noise power spectrum. This may be valid for the central area, for which the hits count map is close to being homogeneous, but it is a poor approximation for the areas outside the central part of the SPDP. It is therefore advisable to include the noise inhomogeneities for the calculation of the residual lensing amplitude. We do this as follows. Let $f_{\text{sky}}(\hat{n}_{\text{lev}})$ be the observed sky fraction with local noise level $\hat{n}_{\text{lev}}$, and $A_{\text{lens}}(\hat{n}_{\text{lev}})$ the predicted residual local lensing amplitude. A weighted average across the SPDP is

$$\frac{1}{f_{\text{sky,tot}}} \int d\hat{n}_{\text{lev}} f_{\text{sky}}(\hat{n}_{\text{lev}}) A_{\text{lens}}(\hat{n}_{\text{lev}}). \quad (57)$$

Here, $f_{\text{sky,tot}}$ is the total observed sky fraction. $f_{\text{sky}}(\hat{n}_{\text{lev}})$ can simply be obtained from the hits count map, and $A_{\text{lens}}(\hat{n}_{\text{lev}})$ is approximated as a polynomial function and calculated from various white-noise and noise-only level predictions. With this, we find that including inhomogeneous noise, $A_{\rm lens}$ increases by about 0.01 at $p_{10}$, and by about 0.03 at $p_{50}$.

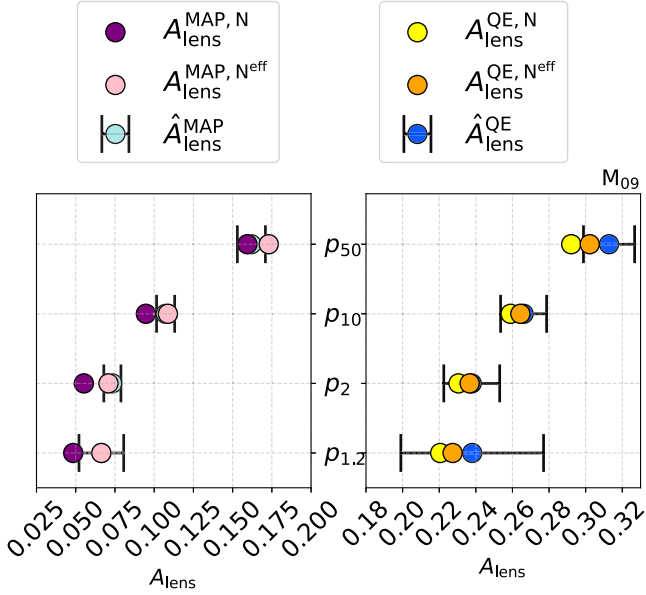

**Figure 12.** Residual lensing amplitude for simulation set $M_{09}$ at the recombination peak, as calculated for different regions inside the observed patch (see Figure 8) from their empirical instrumental noise and foreground residuals. If we include the residual foregrounds as additional noise in our prediction, the iterative prediction increases by a few percent (pink on the left panel), and agrees well with our empirical findings, with the exception of the largest sky region where the noise is highest and least homogeneous, and our prediction scheme is less effective. The impact of the foregrounds on the QE results (right panel) is milder owing to the increased importance of the lensing $B$ power relative to the foreground residuals.

Since $A_{\rm lens}$ is close to linearly dependent on the noise for small $A_{\rm lens}$, we include this correction by simply adding it on top of our predictions for the *noise-only* and *effective noise* case. Figure 12 shows the predictions for $M_{09}$ in purple (noise-only) and pink (effective noise) in the left panel for the MAP solution, and similarly in yellow and orange in the right panel for the QE. The effective noise predictions all seem consistent with our empirical findings, and we see a similar behavior for sky models $M_{00}$ and $M_{09}$ as shown in Figure 11.

Our findings show that for the relevant sky areas, inclusive of the area in which the noise is not too inhomogeneous anymore, the residual foregrounds effectively act as a source of noise in the lensing reconstruction, and simple modeling can accurately predict the reduction in $B$-mode power.

### 4.2.4. Mean Field Tests

The lensing reconstruction likelihood model contains several ingredients, and their accuracy affects the quality of the $B$-lensing template. One factor that relies heavily on good simulations of the data is the mean field, which reflects our (mis-)understanding of induced anisotropies from, among other factors, noise, foreground residuals, and masking. The better we can simulate our noise and foregrounds, the more we can remove their contribution from our estimate. At the same time, it is interesting to understand the impact of having simulations with only limited accuracy. In this section, we discuss the robustness of our lensing reconstruction by altering the mean field, and obtain lensing potential estimates and residual lensing amplitudes that are modified compared to the *baseline* results, which were discussed in the previous section.

Similar to the quadratic estimator mean field, we define the mean field of the iterative solution using an average over a set of converged iterative reconstructions,

$$\hat{\boldsymbol{\alpha}}^{\rm MF} \equiv \langle \hat{\boldsymbol{\alpha}}^{\rm MAP}\rangle_{\rm MC}. \qquad (58)$$

We remind the reader that all lensing estimates $\hat{\boldsymbol{\alpha}}^{\rm MAP}$ are built by subtracting a constant QE mean field at each iteration. Hence, $\hat{\boldsymbol{\alpha}}^{\rm MF}$ is a residual mean field that has not been accounted for by this term: if we could incorporate the mean field term $g_\alpha^{\rm MF}$ from Equation (23) perfectly and at all iterations, then $\hat{\boldsymbol{\alpha}}^{\rm MF}$ defined by Equation (58) would vanish.

Figure 13 shows the lensing mean field deflection field,

$$\hat{\alpha}_{\rm LM}^{\rm MF} \equiv \sqrt{L(L+1)}\hat{\phi}_{\rm LM}^{\rm MF}, \qquad (59)$$

for QE (MAP) in the right (left) column, and for $M_{00}$ (top panels), $M_{07}$ (middle panels), and $M_{09}$ (bottom panels). Figure 14 shows their power spectra in green. In Figure 13, in the QE case, only the largest modes are visible, and are sourced by the mask and noise anisotropies. Their amplitudes are much larger than the expected true lensing signal, and are not visible in the MAP mean field (seen as the wildly different scale), showing that the usage of the approximate QE mean field in the iterations is successful at removing these large signatures. This might be surprising at first since the delensed $B$-mode power is wildly different from the lensed one, which could in principle create large changes in the mean field. However, the $EE$ spectrum is much less affected by delensing, and the mean field is predominantly sourced by the $EE$ quadratic piece. The reason for this is that all mean field sources in these simulations appear as a convergence-like lensing signal rather than shear-like. Small-scale signatures in these maps have large Monte Carlo noise originating from the finite number of simulations (only 100 simulations were used for the tests in this section) used in Equation (58). The $M_{09}$ QE mean field (bottom left panel) has a feature in the lower left area of the patch. A similar structure and a couple of bright spots can also be seen in the MAP case.

The mean field difference maps, between simulation sets (obtained by differencing maps obtained from Equation (59)) are shown in Figure 15. The top two panels show the deflection mean fields of $\hat{\boldsymbol{\alpha}}^{\rm MF}(M_{07}) - \hat{\boldsymbol{\alpha}}^{\rm MF}(M_{00})$, while the bottom two panels show $\hat{\boldsymbol{\alpha}}^{\rm MF}(M_{09}) - \hat{\boldsymbol{\alpha}}^{\rm MF}(M_{00})$. QE is shown in the left panels, and MAP on the right. These differences are ~10% of the total mean field. The top panels show signatures of the foreground amplitude modulation of $fg_{07}$ in the bottom left part of the patch, and in the bottom panels, signatures from foreground filaments of $fg_{09}$ can be seen in the southern part.

In Figure 14, the pink lines show the spectrum of the lensing potential estimates for $M_{00}$ (dotted–dashed), $M_{07}$ (solid), and $M_{09}$ (dashed), $C_L^{\hat{\phi}\hat{\phi}}$, before subtracting their respective mean field. The foregrounds only become relevant in the MAP case: without delensing, the lensing $B$ power dominates the QE reconstruction noise over the foregrounds and all curves are basically indistinguishable.

The calculation of the mean field from an average of simulations like that given in Equation (58) inevitably leads to some level of residual Monte Carlo noise in the estimate due to the finite number of simulations (with expected spectrum $\sim N^{(0)}/N_{\rm MC}$, where $N_{\rm MC}$ is the number of simulations). To avoid this type of noise in the mean field spectra (indicated by

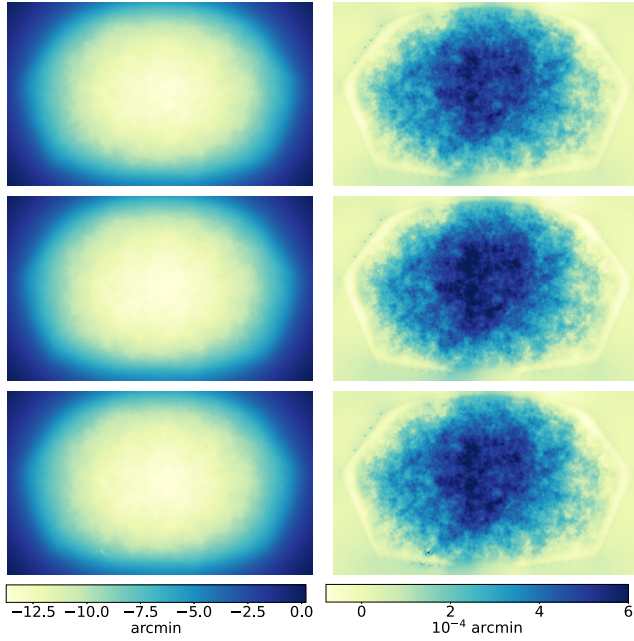

**Figure 13.** QE mean field (left column) and MAP residual mean field (right column) for simulation sets $M_{00}$ (top panels), $M_{07}$ (middle panels), and $M_{09}$ (bottom panels). The MAP mean field is about 3 orders of magnitude smaller than the QE mean field. In fact, most of the small-scale fluctuations are merely noise, but for $M_{09}$ some foreground-induced artifacts are visible in the lower left part of the patch.

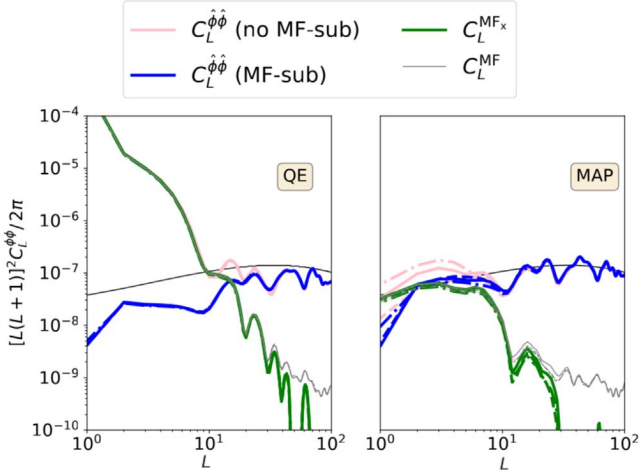

**Figure 14.** Lensing potential estimate pseudo-power spectra for QE (left panel) and MAP (right panel) as obtained from the CMB-S4 LAT simulations. The solid dark line represents the fiducial lensing potential. We show the result for all three simulation sets: $M_{00}$ in dotted–dashed, $M_{07}$ solid, and $M_{09}$ dashed. However, their spectra often overlap and are barely distinguishable, particularly so in the QE case. Pink (blue) shows the QE or MAP lensing potential before (after) mean field subtraction, gray (green) shows the QE mean field or MAP residual mean field with (without) Monte Carlo noise. The mean field drops off quickly for multipoles larger than $L \simeq 10$. The QE mean field (green and gray) contains a large signal at large angular scales, and is the lensing potential's (pink) main component on that scale. On large scales, the residual MAP mean field signal is of the order of the lensing potential, many orders of magnitude smaller than its QE counterpart (green, left panel). This means that at each iteration, the QE mean field is a good proxy for the true mean field term.

the green lines), we have plotted the cross spectrum

$$C_L^{\mathrm{MF}_\times} \equiv \frac{1}{f_{\mathrm{sky}}(2L+1)} \sum_M \hat{\phi}_{LM}^{\mathrm{MF}_1} (\hat{\phi}_{LM}^{\mathrm{MF}_2})^*, \quad (60)$$

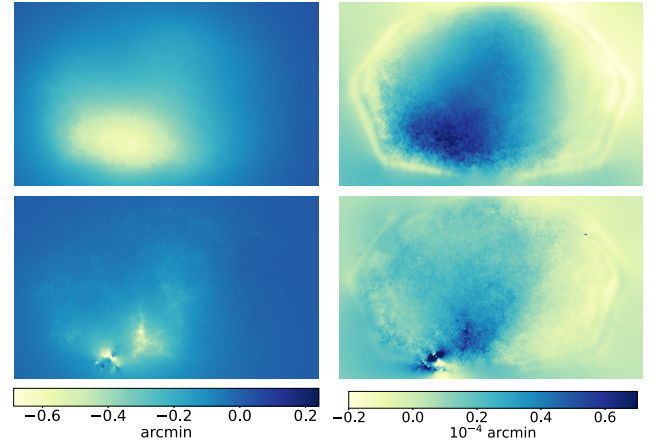

**Figure 15.** Absolute difference maps for the QE (left column) and MAP (right column) deflection mean fields, $\mathrm{MF}_{07} - \mathrm{MF}_{00}$ (top panels), and $\mathrm{MF}_{09} - \mathrm{MF}_{00}$ (bottom panels). There are clear features seen in every panel, which originate at least in part from the foreground residuals.

where the two estimates come from two independent sets of simulations, and we show them in green. Here, $f_{\mathrm{sky}}$ is calculated from the ratio of the number of pixels between the masked and unmasked sky. The Monte Carlo noise enters the iterative reconstruction via the subtracted mean field of the QE, the starting point of the iterations. We show the auto-spectrum of the mean field estimate in gray, and the spectrum of this noise is clearly seen, as the curve departs from the green at higher multipoles. Of course, the mean field template can be set to zero when the noise dominates, and accelerating methods as discussed on the full sky in Section 2.8 can also be used in this setting (though the accuracy of the mean field recovered with these methods relies on that of the likelihood model).

We now turn to three quantitative tests on the impact of the mean field on the delensing efficiency. The changes applied are the following:

(I) A posteriori mean field subtraction: We obtain the average of the MAP reconstructions obtained from Equation (58) and subtract it from each lensing potential estimate (a procedure similar to the QE mean field subtraction).

(II) Swapped mean field: We use the simplest simulation set available (the Gaussian foreground model $M_{00}$) to build the mean field template for the reconstructions on all simulation sets. Hence, the mean field subtraction will not remove the non-Gaussianity of the foregrounds (and also be wrong in a broader sense, since the different models have different power).

(III) Mean $B$-lensing template subtraction (similar to point (I), but at the level of the $B$-lensing templates): we subtract from each $B$ template the average of templates obtained on a set of reconstructions.

In all of these tests, we calculate the residual lensing amplitude on the large, green-contoured sky patch $p_{50}$ (see Figure 8). This includes some mean field signals in the outskirt of the SPDP, yielding conservative results due to the small weight of the SPDP edge for the $r$ inference. For convenience, we only use 100 simulations, with almost exactly the same configuration as described in Section 4.2 (here, we boosted the numerical accuracy parameters to obtain more precise large-scale lenses).

**Table 4**
Mean Field Test Results

| Test | $M_{00}$ | $M_{07}$ | $M_{09}$ |
|---|---|---|---|
| I | $-0.0028 \pm 0.0014$ | $-0.0005 \pm 0.0014$ | $-0.0009 \pm 0.0014$ |
| II | ... | $0.0001 \pm 0.004$ | $0.0039 \pm 0.01$ |
| III | $-0.0005 \pm 0.02$ | $-0.0002 \pm 0.015$ | $-0.0028 \pm 0.03$ |

**Note.** We give the relative change in terms of residual lensing amplitude, compared to the *baseline* configuration, $\frac{\Delta A_{\text{lens}}}{A_{\text{lens}}}$, a negative number indicating an improvement. See the text for the definition of the individual tests.

We obtain $A_{\text{lens}}$ from the ratio of power, following Equation (56), along with the ensemble standard deviation. We summarize our findings in terms of a single value $\frac{\Delta A_{\text{lens}}}{A_{\text{lens}}}$ in Table 4 that compares the tests to the baseline configuration. Here, $\Delta A_{\text{lens}}$ is the difference in residual lensing amplitude between the baseline and modified analyses.

*(I) Mean Field Subtraction.* For test I, we subtract the MAP mean field from the lensing potential estimates prior to the lensing operation, and calculate the $B$-lensing templates, denoted $B_{\text{MF}^-}^{\text{LT}}$. Explicitly,

$$B_{\text{MF}^-}^{\text{LT,MAP}} = (\hat{\alpha}^{\text{MAP}} - \hat{\alpha}^{\text{MF}}) \circ \hat{E}^{\text{unl}}. \quad (61)$$

The result is shown in the first row of Table 4. The impact is always small, and mostly consistent with zero (at $2\sigma$ for simulation set $M_{00}$). The given error bars are the one-simulation equivalents, and therefore, averaging across all simulations, we would detect a tiny improvement in all cases. Nevertheless, this confirms that simply subtracting the QE mean field as calculated for the starting point is sufficient.

*(II) Swapped Mean Field.* We now perform the lensing reconstructions with a mean field different from the baseline. For simulation sets $M_{07}$ and $M_{09}$, we perform reconstructions where the mean field gradient term is set to that of model $M_{00}$ (doing so also changes the starting point for the quadratic estimator but this alone does not affect the converged solution).

Thus, the lensing reconstruction is performed on an inhomogeneous and non-Gaussian foreground simulation set, but its mean field is calculated from a simpler simulation set (with differences shown in Figure 15), with homogeneous and Gaussian foregrounds. The result is shown in Table 4 in the second row, and is consistent with zero. This demonstrates that the choice of mean field in the lensing reconstruction only has a small impact. The mean field from $M_{00}$ only traces the patch mask and noise variance map, yet it nevertheless yields similar results for all simulation sets.

*(III) Mean-B-lensing Template Subtraction.* Figure 16 shows an example $B$-lensing template for $M_{07}$ and $M_{09}$. The MAP mean $B$-lensing templates for $M_{07}$ and $M_{09}$, together with the MAP difference maps, $M_{07} - M_{00}$, and $M_{09} - M_{00}$ are shown in Figure 17. For $M_{09}$ (bottom left panel) we see in Figure 17 clear features along the edge of the patch, hinting at contamination from foreground residuals. $M_{07}$ does not show any dominant features. The difference plots (right panels) show that for $M_{07}$ minus $M_{00}$, the mean $B$-lensing template follows the foreground amplitude modulation of $M_{07}$. For $M_{09}$ minus $M_{00}$, we see bright spots at about the same areas where foreground contamination appears in the mean field of $M_{09}$, also clearly visible in the mean $B$-lensing template in Figure 16. We also see a small amplitude ringing effect along the latitude of the SPDP, which we believe is an artifact of the map-building procedure, introduced when modes below 30 were excised from the maps.

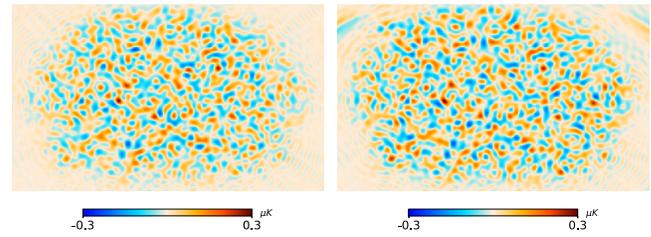

**Figure 16.** $B$-lensing templates for $M_{07}$ (left panel) and $M_{09}$ (right panel). All plots are bandpassed to $30 < \ell < 200$.

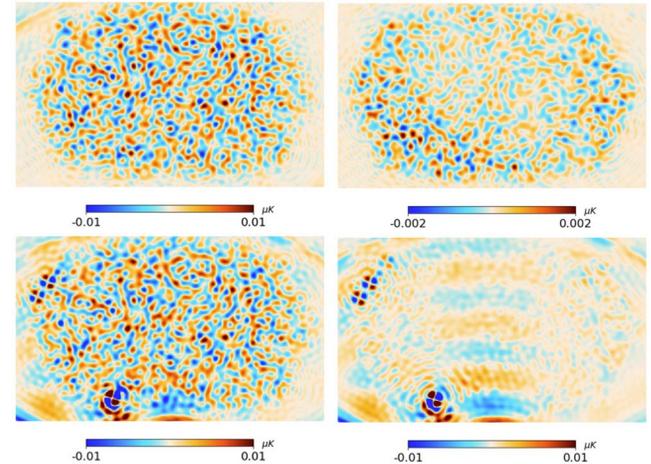

**Figure 17.** Left column: MAP mean $B$-lensing templates for $M_{07}$ (top) and $M_{09}$ (bottom). Right column: difference maps between the mean $B$-lensing templates $M_{07}$ and $M_{09}$ with $M_{00}$. All plots are bandpassed to $30 < \ell < 200$.

We test for the impact of these features by subtracting the mean $B$-lensing template from each $B$-lensing template,

$$B_{\ell m}^{\text{LT,MAP}} \to B_{\ell m}^{\text{LT,MAP}} - \langle B_{\ell m}^{\text{LT,MAP}} \rangle_{\text{MC}}, \quad (62)$$

where in the average on the right-hand side we always exclude the simulation index for which $B^{\text{LT,MAP}}$ is calculated.

However, similar to the QE and MAP mean field, subtracting not only removes contaminants, but also adds some level of Monte Carlo noise. For this reason, we calculate power spectra by dividing the simulation sets into two splits and taking cross spectra. The result is shown in the third column of Table 4. We do not find any improvement, and the error bars are rather large.

Summarizing these tests, they confirm that our results are robust against changes to the mean field treatment, and more generally that the impact of the mean field is small.

### 4.3. Chile Deep Patch Delensing

As a last test, we repeat the calculation of the $B$-lensing template for a different CMB-S4 configuration. Specifically, we assume that the Deep Field would be observed from Chile, referring to this as the Chile Deep Patch (CDP). Observing the sky from Chile would, among other things, change the scan strategy of the telescopes, therefore impacting the noise levels and observed sky fractions. Figure 18 shows the derived noise-level map after component separation, in equatorial coordinates. The central noise level is $\hat{n}_{\text{lev}} = 0.6 \, \mu\text{K}$ arcmin and therefore about 50% higher than for SPDP, and the scan area is much broader.

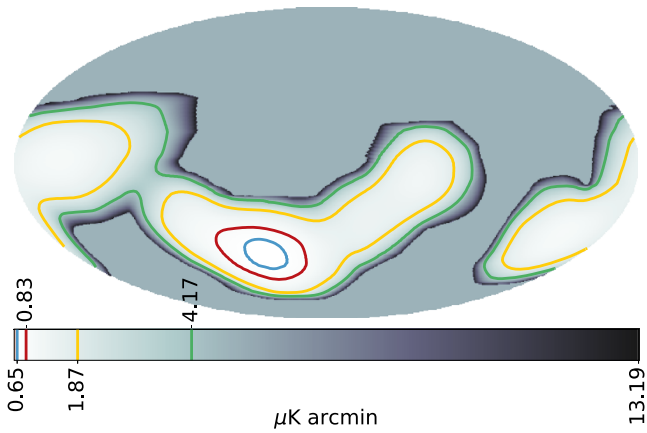

**Figure 18.** This is a full sky noise-level map for the Chile configuration and in equatorial coordinates. The contours show edges that contain the masks A (blue), B (red), C (green), and D (yellow), and the numbers in the color bar refer to the contour noise level. The lowest (highest) noise of about 0.65 (13.19) $\mu$K arcmin is indicated by the blue (dark gray) contour and all are derived from the Chile hits count map.

We generate two simulation sets, that is, the foreground models $M_{00}$ and $M_{07}$, and repeat lensing reconstruction for 100 realizations of the CDP component-separated maps ($M_{09}$ was not available over the larger area at this time owing to practical difficulties). We update our noise model and overlapping $B$-mode deprojection matrix to this case, and leave other ingredients of our algorithm essentially the same.

To improve numerical stability for the iterative lensing reconstruction step, we have increased the mask by excluding areas for which the noise is more than 100 times larger than the central noise value. These are mostly pixels near the galactic plane with high foreground residuals that can safely be discarded. Iterative lensing reconstruction succeeds in the entire resulting area.

The bulk of the constraining power comes from the red region in Figure 18. The larger sky patches indicated by the yellow and green contours partly cover the Galactic plane. Although lensing reconstruction does converge there as well, the foreground residuals are too large in these areas, leading to uninteresting delensing results.

In the red region and for $M_{07}$, we find[35] $A_{\text{lens}}^{\text{QE}} = 0.30 \pm 0.02$ and $A_{\text{lens}}^{\text{MAP}} = 0.121 \pm 0.007$. For $M_{00}$, we find $A_{\text{lens}}^{\text{QE}} = 0.29 \pm 0.02$ and $A_{\text{lens}}^{\text{MAP}} = 0.115 \pm 0.006$.

These empirical results for iterative delensing are in reasonable agreement with our theoretical predictions, $A_{\text{lens}}^{\text{MAP}} = 0.11$ for $M_{07}$, and $A_{\text{lens}}^{\text{MAP}} = 0.10$ for $M_{00}$, matching as well those expected from the CMB-S4 $r$-forecast paper (Abazajian et al. 2022).

### 4.4. Constraints on r

In our previous CMB-S4 $r$-constraint paper (Abazajian et al. 2022) one of the two reanalysis methods used was the parametric multicomponent cross-spectral likelihood that was first introduced in BICEP2/Keck Collaboration et al. (2015), and which has been used in all BICEP/Keck analyses up to the most recent (Ade et al. 2021). Here, we take this likelihood and add the lensing templates derived in the previous section as an additional pseudo-frequency band, in a similar manner to the joint analysis of BICEP/Keck and SPT data presented in BICEP/Keck Collaboration et al. (2021).

---

[35] The results for QE are calculated using nonperturbative remapping of the $E$ mode, improving performance slightly, as before.

To obtain sufficient $E/B$ separation purity we calculate the auto and cross spectra between the maps using the S²HAT library (Grain et al. 2009). The resulting band powers are then compared against a likelihood model built using the Hamimeche-Lewis (Hamimeche & Lewis 2008) approximation with the following parameters: the tensor-to-scalar ratio $r$; the amplitudes of the dust and synchrotron $B$-mode power spectra, their frequency spectral indices, and their spatial power law indices; the frequency-independent spatial correlation between dust and synchrotron; and the dust correlation between 217 and 353 GHz. See Appendix G of BICEP2 Collaboration et al. (2018) for a full description of the parameterization.

The likelihood requires statistical characterization of the signal and noise parts of the lensing template auto-spectrum and of their cross spectra to the SAT frequency maps. In this first analysis, we do this in the following crude (but effective) manner. Making the assumption that the template may be written as a filtered version of the true CMB lensing $B$ mode plus an additive noise term, we compute an effective, isotropic filter function from 100 cross spectra to their known simulation CMB inputs (neglecting the precise position dependence of the filter). This allows us to build predictions for the cross spectra to the SAT maps. Further, we obtain the lensing template auto-spectrum noise power by subtracting from the templates spectra the signal term as predicted from the isotropic filter model.

For each of the 500 simulation realizations, and each of the three foreground models, we perform a maximum likelihood (ML) search, both with and without the lensing template. We use the mean and the standard deviation of a given parameter value to assess the bias and experimental uncertainty of that parameter—where the parameter $r$ is of course by far of the greatest interest. The SPDP results are shown in Figure 19 for the realizations with $r = 0$ and those with $r = 0.003$. As expected, the addition of the lensing template (effective delensing) dramatically reduces the uncertainty $\sigma$ on $r$, here denoted $\sigma(r)$. In the case of the non-Gaussian model $M_{09}$, a significant bias is also removed. Table 5 gives the with-template results in numerical form. These can be compared to the results in the rightmost column of Table 2 of our previous paper (Abazajian et al. 2022). The results for $M_{00}$ are most directly comparable. Previously for $r = 0$ we had $\sigma(r) = 5.7 \times 10^{-4}$, to be compared to the present $4.3 \times 10^{-4}$, the latter being below the $\sigma(r) = 5 \times 10^{-4}$ science goal of CMB-S4 (Abazajian et al. 2019).

We have preliminary ML search runs for the CDP simulations as well. The uniform amplitude $M_{00}$ model is a gross oversimplification for this larger sky patch. Nevertheless, for input $r = 0$ this model gives $\sigma(r) = 5.2 \times 10^{-4}$, which is consistent with the expectation that the best sensitivity will be obtained with the most concentrated sky coverage. For $M_{07}$ the effective foreground amplitude averaged over the patch is extremely high and the likelihood results are not robust. We find that $\sigma(r)$ is only increased to $7.0 \times 10^{-4}$, but there is significant bias—further investigations are ongoing. No results are available for $M_{09}$ at the time of this writing.

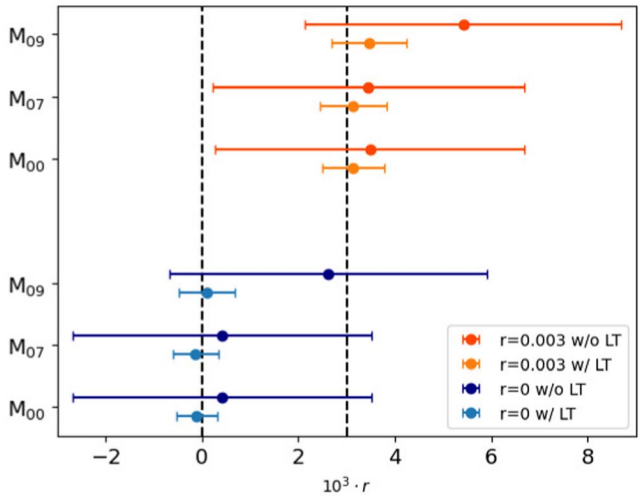

**Figure 19.** Results from reanalysis of map-based simulations of CMB-S4 observing the SPDP. The mean and standard deviation of the ML $r$ values evaluated over ensembles of simulation realizations are shown as the central point and horizontal error bar, respectively. Blue and orange points show the results for the $r = 0$ and $r = 0.003$ realizations, with the vertical lines indicating these values. The darker points are the results without the lensing template, while the lighter points show the substantial benefit of including the iteratively reconstructed templates described in this paper.

**Table 5**
Results from the Reanalysis of Map-based Simulations of CMB-S4 Observing the SPDP, Including the Reconstructed Lensing Templates Described in This Paper

| $r$ Value | Sky Model | $\sigma(r) \times 10^4$ | $r$ Bias $\times 10^4$ |
|---|---|---|---|
| 0 | 0 | 4.3 | −1.0 |
|   | 7 | 4.7 | −1.3 |
|   | 9 | 5.8 | 1.1 |
| 0.003 | 0 | 6.4 | 1.3 |
|   | 7 | 6.9 | 1.4 |
|   | 9 | 7.8 | 4.7 |

## 5. Conclusion

In this paper, we have presented an optimal lensing reconstruction software on the curved sky, usable under realistic conditions, inclusive of inhomogeneous noise, and on sky patches of arbitrary size. With it, we achieve the CMB-S4 science goal on $r$ by removing 92%–93% of the $B$-lensing power, depending on the simulation set.

The method mostly generalizes that of Carron & Lewis (2017) to curved sky geometry. We discussed carefully all ingredients, and showed that the few approximations that are made are perfectly under control given the moderate size of $\Lambda$CDM deflection angles (as is often the case, the sky curvature can be neglected on the scale of the deflection), and that the reconstruction should be optimal provided that the CMB likelihood model is a fair representation of the data.

We found most difficulties of a practical nature in the reconstruction process at low-lensing multipoles. There, the mean field is prominent, and more generally, the behavior of the reconstruction of the largest scale lenses is less stable, and also often requires higher accuracy in the most important steps (lensing remappings, recovery of the delensed $E$ mode) than for smaller-scale lenses. We have used here as the baseline a compromise, where the mean field and largest lenses are not precisely reconstructed, which for most applications is perfectly satisfactory: these large lenses contribute little to $B$-mode delensing on degree scales, and for the purpose of lensing spectrum reconstruction, cosmic variance dominates there anyway, so that the quadratic estimator solution can be satisfactory. Recent work (Reinecke et al. 2023) has allowed us to solve most of these issues, making upcoming analyses easier as well as slightly improving the lensing reconstruction further.

New to this paper is the characterization of the noise anisotropies induced by delensing as the only source of the mean field, which has no direct analog in quadratic estimator theory. We showed that it is a negligible source of complication, in particular, for situations where uncertainties in the statistical properties of the data overshadow its contribution. One naturally leads to drastic simplifications in the treatment of the mean field, which becomes as simple as for the standard quadratic estimator.

The iterative reconstruction usually converges in a reasonable number of steps (rarely did we use more than 15 iterations), but is by no means a smooth process. In fact, the non-Gaussian nature of the likelihood is such that for low-noise levels full Newton steps are much too strong, resulting generically in unrealistic, noninvertible deflection field estimates. We found that reducing the step in a scale-dependent manner can correct that problem, but this requires undesirable ad hoc tuning, with little improvements each time, except for the first few iterations; calculation of the gradients is costly, and generally, improvement of the convergence after a few steps would be highly desirable and is left for future work. For the purpose of delensing, it helps that recovery of the lensing $B$ mode appears to be easier than that of the lensing deflection field and $E$-mode map individually: errors in both tracers display statistical dependence in order to still produce the right level of $B$ modes seen in the data.

Also new to this paper is the joint reconstruction of the lensing post-Born curl together with the standard lensing gradient mode. The joint reconstruction brought no complications with it, and the output is correctly described by our predictions of the iterative $N^{(0)}$ and $N^{(1)}$ biases. At the current time, the lensing curl is really only useful as a sanity check of the lensing reconstruction, but at the CMB-S4 depth probed here, it will also be a signal, whose detectability is boosted by large factors with our optimal method.

We focused mostly on simulations of a very deep patch, using semi-realistic inhomogeneous CMB-S4 noise maps, together with two non-Gaussian foreground models. Our baseline reconstructions on ILC-cleaned maps are not perfect near the edges and on large scales. Nevertheless, this has little impact overall, and the tracer is able to remove close to 93% of the lensing over the relevant area, as expected. Of course, available Galactic foreground models are still crude on small scales, but we find encouragingly that the main effect of the presence of the foregrounds on the lensing templates appears to be the increased effective noise of the maps. This extends the cautious optimism of Beck et al. (2020) to the case of iterative delensing. For the most complex model $M_{09}$ a non-Gaussian feature is present close to the edges, which is visible when averaging the templates.

We built predictions for the delensing capabilities and lensing bias calculations, using a slightly modified version of the well-known algorithm introduced in Smith et al. (2012). It is quite noteworthy how well these predictions (which do not exactly follow from first principles) seem to work even under nonidealized conditions, and it is reassuring to see that forecasts with these tools can probably be trusted.

To summarize, Kamionkowski & Kovetz (2016) write that *B*-mode delensing is an "ambitious, sophisticated, and challenging endeavor": we believe that this paper makes the prospect of an eventually successful delensing much more realistic. Of course, much work remains to be done. Our noise simulations were inhomogeneous but otherwise did not have the full complexity expected from a ground-based experiment like CMB-S4. While at least some of such nonidealities are expected to impact predominantly scales larger than those most relevant for lensing reconstruction, it is difficult to assess the importance of these effects for delensing without more realistic simulations. This is left for future work. Other areas that need development include the study of systematic effects, better foreground models and cleaning methods, nonlinear large-scale structure simulation inputs, and further code improvements.

## Acknowledgments

The authors thank Victor Buza and Benjamin Racine for previous work relevant to this paper. CMB-S4 is supported by the Director, Office of Science, Office of High Energy Physics of the U.S. Department of Energy under contract No. DE-AC02-05CH11231; by the National Energy Research Scientific Computing Center, a DOE Office of Science User Facility under the same contract; and by the Divisions of Physics and Astronomical Sciences and the Office of Polar Programs of the U.S. National Science Foundation under Mid-Scale Research Infrastructure award OPP-1935892. Considerable additional support is provided by the many CMB-S4 team members and their institutions. This research used resources of the National Energy Research Scientific Computing Center, a DOE Office of Science User Facility supported by the Office of Science of the U.S. Department of Energy under contract No. DE-AC02-05CH11231. J.C., S.B., and L.L. acknowledge support from a SNSF Eccellenza Professorial Fellowship (No. 186879). C.B. acknowledges support from the U.S. Department of Energy Office of Science (DE-SC0011784). This work was supported by a grant from the Swiss National Supercomputing Center (CSCS) under project IDs s1203 and sm80. Some computations in this paper were run on the Odyssey/Cannon cluster supported by the FAS Science Division Research Computing Group at Harvard University.

## Appendix A
## Spin-weighted Fields and Harmonics

Throughout this paper, we make heavy use of spin-weighted spherical harmonic (Newman & Penrose 1966; Goldberg et al. 1967; Lewis et al. 2002) transforms with spins between 0 and 3. A spin-weighted field $_sf(\hat{n})$ ($s \geq 0$) on the sphere and parametrized with colatitude and longitude $(\theta, \varphi)$ is defined with reference to the local axes, $\boldsymbol{e}_\theta$ and $\boldsymbol{e}_\varphi$. Our conventions are that $\boldsymbol{e}_\theta$ and $\boldsymbol{e}_\varphi$ point southward and eastward, respectively.

Under a clockwise rotation at $\hat{n}$ by some angle $\psi$, the field transforms, by definition, is

$$_sf(\hat{n}) \rightarrow e^{is\psi(\hat{n})} \, _sf(\hat{n}). \quad (A1)$$

A spin-*s* field can conveniently be decomposed into its gradient (*G*) and curl (*C*) harmonics by using the spin-weighted spherical harmonics, $_sY_{\ell m}(\hat{n})$. The expansion has the form

$$_{\pm s}f(\hat{n}) = -(\pm 1)^s \sum_{\ell m} (G_{\ell m} \pm i C_{\ell m}) \, _{\pm s}Y_{\ell m}(\hat{n}). \quad (A2)$$

Further, let $\eth^+$ and $\eth^-$ be the spin-raising and spin-lowering operators (Newman & Penrose 1966; Goldberg et al. 1967),

$$\eth_s^\pm f = -(\sin\theta)^{\pm s}\left(\partial_\theta \pm \frac{i}{\sin\theta}\partial_\varphi\right)(\sin\theta)^{\mp s} \, _s f. \quad (A3)$$

We follow generally the conventions of Lewis et al. (2002), and refer to the appendices of that paper for a discussion in the context of the CMB. With the help of these operators, one can introduce orthonormal basis functions (the spin-weighted spherical harmonics $_sY_{\ell m}$) for spin-weighted fields. Explicitly,

$$\eth_s^\pm Y_{\ell m} = \pm\sqrt{(\ell \mp s)(\ell \pm s + 1)} \, _{s\pm 1}Y_{\ell m}. \quad (A4)$$

A lensing remapping on the sphere can be described by a vector field (concretely, a field of spin-1), and this vector field can, in turn, be described by two scalar (spin-0) lensing potentials, $\phi(\hat{n})$ and $\Omega(\hat{n})$, with the spin-1 deflection vector defined as follows:

$$\begin{aligned}
_1\alpha(\hat{n}) &= -\bar{\eth}^+(\phi(\hat{n}) + i\,\Omega(\hat{n})) \\
&= -\sum_{\ell m}\sqrt{\ell(\ell+1)}(\phi_{\ell m} + i\Omega_{\ell m}) \, _1Y_{\ell m}(\hat{n}).
\end{aligned} \quad (A5)$$

The second line follows from Equation (A4). As can be seen by comparing this to Equation (A2), $\phi$ defines the gradient and $\Omega$ the curl component of the deflection vector. For a pure gradient deflection field, its real and imaginary components at $\hat{n}$ are simply $\partial_\theta \phi(\hat{n})$ and $\csc(\theta)\partial_\varphi \phi(\hat{n})$, and are generally denoted $\alpha_\theta$ and $\alpha_\varphi$. If convenient, we also make use of the polar representation of the vector, with $\alpha$ the magnitude of the field, and $\beta$ the angle to the $\theta$ coordinate direction, such that, at each point on the sphere, the following relation holds:

$$_{\pm 1}\alpha(\hat{n}) \equiv \alpha_\theta(\hat{n}) \pm i\alpha_\varphi(\hat{n}) \equiv \alpha(\hat{n})e^{\pm i\beta(\hat{n})}. \quad (A6)$$

## Appendix B
## Delensed Noise Mean Field

This section gives details on the delensed noise mean field (the variation of the covariance matrix log-determinant with $\boldsymbol{\alpha}$) in the case of an otherwise isotropic configuration. Using the Woodbury determinant relation, the relevant log-determinant is

$$\frac{1}{2}\ln\det[C^{EE,\text{unl},-1} + N_\alpha^{-1}] \equiv \frac{1}{2}\ln D, \quad (B1)$$

with the inverse delensed *E*-mode noise matrix

$$N_\alpha^{-1} = \, _2\mathcal{Y}^\dagger \mathcal{D}_\alpha^\dagger \mathcal{B}^\dagger N^{-1} \mathcal{B} \mathcal{D}_\alpha \, _2\mathcal{Y}. \quad (B2)$$

There are two variations to be considered. The first variation of the log-determinant can be written in the following way, using simple matrix algebra:

$$\begin{aligned}
\delta \ln D &= \text{Tr}\, C^{EE,\text{unl}}(C^{EE,\text{unl}} + N_\alpha)^{-1}\delta \ln N_\alpha^{-1} \\
&\equiv \text{Tr}\, \mathcal{W}_\alpha \delta \ln N_\alpha^{-1}.
\end{aligned} \quad (B3)$$

For convenience, we have used the notation $\delta \ln N_\alpha^{-1}$ for the matrix $N_\alpha \delta N_\alpha^{-1}$. This matrix has roughly constant contributions on all scales. It is contracted with the Wiener-filtering matrix (defined as $\mathcal{W}_\alpha$ in the second line), thereby suppressing contributions from noisy $E$ modes. The impact of $\alpha$ in the Wiener filter matrix is, however, only significant when the noise is comparable to $C^{EE,\mathrm{unl}}$. Hence, for low-noise experiments, the isotropic Wiener filter $\mathcal{W}_\ell = C_\ell/(C_\ell + N_\ell)$, which is the leading-order contribution, already gives an almost exact result. We now discuss $\delta \ln N_\alpha^{-1}$ in more detail, and then look at the second variation as well.

The action of lenses much larger than the beam commutes with the beam to good accuracy. It is useful to consider this approximation first, which results in the correct low-$L$ behavior, and which also renders the results more transparent. This assumption also corresponds to the exact result in the case of zero beamwidth (but arbitrarily large white-noise level, so that this still corresponds to a somewhat physically relevant situation). Under this assumption, we can swap the order of appearance of the beam and deflection operations in Equation (B2). Assuming homogeneous noise with full sky coverage, and using the exact relation $[\mathcal{D}_\alpha^\dagger \mathcal{D}_\alpha][\hat{n}, \hat{n}'] = |A_{\alpha^{-1}}|(\hat{n})\delta^D(\hat{n}, \hat{n}')$, the noise matrix greatly simplifies.

For convenience, we now adopt a notation convenient for intensity measurements by using spin-0 spherical harmonics. However, the simulations discussed in this paper include polarized data and have no primordial $B$ modes, and we generally must use spin-2 harmonics to correctly describe it. These two differences lead to slight complications in the argument below. They are, nevertheless, only formal, and lead to identical leading-order and similar second-order results, and will be discussed at the end.

The inverse-noise matrix, in the approximation discussed above, becomes

$$[N_\alpha^{-1}]_{\ell m, \ell' m'} = \int d\hat{n} \frac{|A_{\alpha^{-1}}|(\hat{n})}{N_\ell^{1/2} N_{\ell'}^{1/2}} Y_{\ell m}^\dagger(\hat{n}) Y_{\ell' m'}(\hat{n}). \quad \text{(B4)}$$

Assuming that the band limits give a sufficiently large range, the spherical harmonics project onto a complete set, and the matrix has the explicit inverse

$$[N_\alpha]_{\ell m, \ell' m'} = \int d^2\hat{n} \frac{N_\ell^{1/2} N_{\ell'}^{1/2}}{|A_{\alpha^{-1}}|(\hat{n})} Y_{\ell m}^\dagger(\hat{n}) Y_{\ell' m'}(\hat{n}). \quad \text{(B5)}$$

With this, we find that the variation $N_\alpha \delta N_\alpha^{-1}$ in Equation (B3) is

$$[\delta \ln N_\alpha^{-1}]_{\ell m, \ell' m'}$$
$$= \frac{N_\ell^{1/2}}{N_{\ell'}^{1/2}} \int d^2\hat{n}\; \delta \ln |A_{\alpha^{-1}}|(\hat{n}) Y_{\ell m}^\dagger(\hat{n}) Y_{\ell' m'}(\hat{n}). \quad \text{(B6)}$$

Further, we can then write the leading term of Equation (B3) as

$$\delta \ln D = \left(\delta \int \frac{d^2\hat{n}'}{4\pi} \ln A_{\alpha^{-1}}(\hat{n}')\right)\left(\sum_\ell (2\ell+1)\mathcal{W}_\ell\right). \quad \text{(B7)}$$

By using $\ln A_{\alpha^{-1}}(\hat{n}') = -\ln A_\alpha(\hat{n})$, where $\hat{n}'$ and $\hat{n}$ are the deflected and undeflected positions, respectively, and performing the change of parameterization in the integral, the prefactor is

$$-\delta \int \frac{d^2\hat{n}}{4\pi} |A_\alpha|(\hat{n}) \ln |A_\alpha|(\hat{n}) \sim -\delta \int \frac{d^2\hat{n}}{4\pi} 2\kappa^2(\hat{n}). \quad \text{(B8)}$$

By definition, $g_{\mathrm{LM}}^{\mathrm{MF},\kappa} = \frac{1}{2}\delta \ln D / \delta \kappa_{\mathrm{LM}}$, and we recover

$$g_{\mathrm{LM}}^{\mathrm{MF},\kappa} = -2\kappa_{\mathrm{LM}}\left(\sum_\ell \frac{2\ell+1}{4\pi}\mathcal{W}_\ell\right), \quad \text{(B9)}$$

as given in the text in Equation (24).

Let us now discuss the second variation. It is natural to continue working with $\ln A_{\alpha^{-1}}$, since the second variation $\delta^2 \ln N_\alpha^{-1}$ vanishes (see Equation (B6)). For this reason, the only contribution is from the change of the Wiener filter. This may be written as

$$\delta^2 \ln D = [\mathrm{Tr}\; \mathcal{W}_\alpha \delta \ln N_\alpha^{-1}(1-\mathcal{W}_\alpha)\delta \ln N_\alpha^{-1}]. \quad \text{(B10)}$$

The full result following from this is given in Equation (B12). The squeezed limit, however, (low $L$, high $\ell$, and $\ell \sim \ell'$) is easily extracted:

$$\left(\int \frac{d^2\hat{n}}{4\pi} |A_\alpha|(\ln|A_\alpha|)^2\right)\left(\sum_\ell (2\ell+1)\mathcal{W}_\ell(1-\mathcal{W}_\ell)\right). \quad \text{(B11)}$$

The prefactor is $\int (2\kappa)^2 +$ higher orders. This is of the same order in $\kappa$ as the first variation, but with much lower relevance due to $\mathcal{W}_\ell(1-\mathcal{W}_\ell)$ and taking only contributions when $C_\ell \sim N_\ell$. The exact second variation result for spin-0 intensity is

$$\frac{1}{2}\ln D \ni \frac{1}{2}\sum_{\mathrm{LM}}|2\kappa_{\mathrm{LM}}|^2$$
$$\cdot 2\pi \int_{-1}^1 d\mu\; d_{00}^L(\mu)\xi_{00}^{\mathcal{W}}(\mu)\xi_{00}^{(1-\mathcal{W})}(\mu), \quad \text{(B12)}$$

with for any spins $a$, $b$, and filter $F$,

$$\xi_{ab}^F(\mu) = \sum_\ell \frac{2\ell+1}{4\pi} F_\ell d_{ab}^\ell(\mu). \quad \text{(B13)}$$

For polarization, and in the limit of vanishing $C_\ell^{BB,\mathrm{unl}}$, the differences are the following. First, the Wiener filter $\mathcal{W}_\alpha$ becomes pure $EE$ to leading order, but $(1-\mathcal{W}_\alpha)$ has both an $EE$ and a $BB$ component. This last term originates from the contribution of the $EB$ part of the quadratic estimator. Second, we must use the spin-2 harmonics. The first variation is unchanged, but the second becomes

$$\frac{1}{2}\ln D \ni \frac{1}{2}\sum_{\mathrm{LM}}|2\kappa_{\mathrm{LM}}|^2$$
$$\times \Bigg\{2\pi \int_{-1}^1 d_{00}^L(\mu) \frac{1}{2}[\xi_{2,2}^{\mathcal{W}}(\mu)\xi_{-2,-2}^{1-\mathcal{W}}(\mu) + \xi_{-2,2}^{\mathcal{W}}(\mu)\xi_{2,-2}^{1-\mathcal{W}}(\mu)]$$
$$+2\pi \int_{-1}^1 d_{00}^L(\mu)\frac{1}{2}[\xi_{2,2}^{\mathcal{W}}(\mu)\xi_{-2,-2}^1(\mu) - \xi_{-2,2}^{\mathcal{W}}(\mu)\xi_{2,-2}^1(\mu)]\Bigg\},$$

where the second line is the $EB$ contribution. Combining this with Equation (B9) gives the final result for the mean field for tiny beams,

$$g_{LM}^{MF,\kappa} = -2\kappa_{LM}\Bigg\{\left[\sum_\ell \frac{2\ell+1}{4\pi}\mathcal{W}_\ell\right]$$
$$-2\pi\int_{-1}^1 d\mu\, d_{00}^L(\mu)\frac{1}{2}[\xi_{2,2}^\mathcal{W}(\mu)\xi_{-2,-2}^{1-\mathcal{W}}(\mu) + \xi_{-2,2}^\mathcal{W}(\mu)\xi_{2,-2}^{1-\mathcal{W}}(\mu)]$$
$$-2\pi\int_{-1}^1 d\mu\, d_{00}^L(\mu)\frac{1}{2}[\xi_{2,2}^\mathcal{W}(\mu)\xi_{-2,-2}^1(\mu) - \xi_{-2,2}^\mathcal{W}(\mu)\xi_{2,-2}^1(\mu)]\Bigg\}.$$
(B14)

The first two lines are the leading and sub-leading $EE$ contributions, while the third line is the one from the $EB$ part of the quadratic estimator, which is basically zero at low $L$, as expected.

## Appendix C
## Magnification Determinant

In this section, we describe a series of steps to derive the exact form of the remapping Jacobian determinant in Equation (15).

In standard polar coordinates, we use the local orthonormal basis at $\hat{n}$, yielding $\boldsymbol{e}_r = \hat{n}$, $\boldsymbol{e}_\theta$, and $\boldsymbol{e}_\varphi$. Variations along $\boldsymbol{e}_\theta$ and $\boldsymbol{e}_\varphi$ induce at the deflected position $\hat{n}'$ shifts $\partial_\theta \hat{n}'$ and $\frac{1}{\sin\theta}\partial_\varphi \hat{n}'$ in the plane tangent to the sphere at $\hat{n}'$. We can write these shifts using $\boldsymbol{e}_r$, $\boldsymbol{e}_\theta$, and $\boldsymbol{e}_\varphi$.

With this, the Jacobian determinant can be written as

$$|A| = |\partial_\theta \hat{n}' \times \partial_{\tilde{\varphi}} \hat{n}'|, \tag{C1}$$

where we have used the notation $\partial_{\tilde{\varphi}}$ for $\frac{1}{\sin\theta}\partial_\varphi$. The third deflection vector $\boldsymbol{d} = \hat{n}' - \hat{n}$ can be written on this basis as

$$\hat{n}' = \hat{n} + \boldsymbol{d} \equiv (1 + \tilde{\alpha}_r)\boldsymbol{e}_r + \tilde{\alpha}_\theta \boldsymbol{e}_\theta + \tilde{\alpha}_\varphi \boldsymbol{e}_\varphi, \tag{C2}$$

with

$$1 + \tilde{\alpha}_r = \cos\alpha, \quad \tilde{\alpha}_\theta = \frac{\sin\alpha}{\alpha}\alpha_\theta, \quad \tilde{\alpha}_\varphi = \frac{\sin\alpha}{\alpha}\alpha_\varphi. \tag{C3}$$

The variation of $\hat{n}'$ can be calculated from the variations of $_1\tilde{\alpha}$, and those of $\boldsymbol{e}_r$, $\boldsymbol{e}_\theta$, and $\boldsymbol{e}_\varphi$. A straightforward, albeit somewhat lengthy calculation, gives

$$\partial_\theta \hat{n}' \times \partial_{\tilde{\varphi}} \hat{n}'$$
$$= \begin{pmatrix} \partial_\theta \tilde{\alpha}_r - \tilde{\alpha}_\theta \\ 1 + \tilde{\alpha}_r + \partial_\theta \tilde{\alpha}_\theta \\ \partial_\theta \tilde{\alpha}_\varphi \end{pmatrix} \times \begin{pmatrix} \partial_{\tilde{\varphi}}\tilde{\alpha}_r - \tilde{\alpha}_\varphi \\ \partial_{\tilde{\varphi}}\tilde{\alpha}_\theta - \tilde{\alpha}_\varphi \cot\theta \\ 1 + \tilde{\alpha}_r + \tilde{\alpha}_\theta \cot\theta + \partial_{\tilde{\varphi}}\tilde{\alpha}_\varphi \end{pmatrix}. \tag{C4}$$

With the definitions of the convergence and field rotation, $\tilde{\kappa} + i\tilde{\omega} = \frac{1}{2}\bar{\eth}\,_1\tilde{\alpha}$, shear $\tilde{\gamma} = \frac{1}{2}\eth\,_1\tilde{\alpha}$, as well as a new spin-1 field $_1f \equiv f_1 + if_2 \equiv {}_1\tilde{\alpha} + \eth\tilde{\alpha}_r$, we can write

$$\partial_\theta \hat{n}' \times \partial_{\tilde{\varphi}} \hat{n}' = \begin{pmatrix} (1 + \tilde{\alpha}_r - \tilde{\kappa})^2 + \tilde{\omega}^2 - \tilde{\gamma}^2 \\ -f_1(1+\tilde{\alpha}_r - \tilde{\kappa} + \tilde{\gamma}_1) + f_2(-\tilde{\gamma}_2 - \tilde{\omega}) \\ -f_1(-\tilde{\gamma}_2 + \tilde{\omega}) + f_2(1 + \tilde{\alpha}_r - \tilde{\kappa} - \tilde{\gamma}_1) \end{pmatrix}. \tag{C5}$$

The first ($\boldsymbol{e}_r$) component reduces to the usual magnification matrix determinant $1 - 2\kappa$ for infinitesimal deflections. The $\boldsymbol{e}_\theta$ and $\boldsymbol{e}_\varphi$ components form the real and imaginary parts of a spin-1 field, giving a correction quadratic in the deflection angle.

To simplify all the calculations, we introduce and calculate with spin-0 quantities $\tilde{\eta}, \eta$ and $\tilde{\xi}, \xi$, built from the three-dimensional and tangential deflection, respectively,

$$\tilde{\eta} \equiv 1 + \tilde{\alpha}_r - \tilde{\kappa} + i\tilde{\omega}, \quad \eta \equiv 1 - \kappa + i\omega,$$
$$\tilde{\xi} \equiv e^{-2i\beta}\tilde{\gamma}, \quad \xi \equiv e^{-2i\beta}\gamma. \tag{C6}$$

A short calculation, applying $\eth$ on $\tilde{\eta}$ and $\tilde{\xi}$ shows that they are quite simply related to their tangential counterparts:

$$\begin{pmatrix}\tilde{\eta}\\\tilde{\xi}\end{pmatrix} = \begin{pmatrix} j_0(\alpha) - \frac{\alpha}{2}j_1(\alpha) & \frac{\alpha}{2}j_1(\alpha) \\ \frac{\alpha}{2}j_1(\alpha) & j_0 - \frac{\alpha}{2}j_1(\alpha) \end{pmatrix}\begin{pmatrix}\eta\\\xi\end{pmatrix}$$
$$- \frac{\alpha}{2}j_1(\alpha)\begin{pmatrix}1\\1\end{pmatrix}, \tag{C7}$$

with $j_0(\alpha) = \sin(\alpha)/\alpha$ and $\alpha j_1(\alpha) = -\cos(\alpha) + \sin(\alpha)/\alpha$ the first two spherical Bessel functions. With this notation in place, the $\boldsymbol{e}_r$ component of the vector product is

$$(1 + \tilde{\alpha}_r - \tilde{\kappa})^2 + \tilde{\omega}^2 - \tilde{\gamma}^2 = |\tilde{\eta}|^2 - |\tilde{\xi}|^2$$
$$= \cos\alpha\left(j_0(\alpha)(|\eta|^2 - |\xi|^2) - \frac{\alpha}{2}j_1(\alpha)(\eta + \eta^* - \xi - \xi^*)\right). \tag{C8}$$

The contributions of the $\boldsymbol{e}_\theta$ and $\boldsymbol{e}_\varphi$ can also be expressed easily in terms of $\eta$ and $\xi$, with the exact same result as Equation (C8), but with a $\sin\alpha$ prefactor replaced by $\cos\alpha$. The full Jacobian determinant then collects a factor of $\cos^2\alpha$ and $\sin^2\alpha$ from the component parallel and perpendicular to $\hat{n}$, respectively, with the result

$$|A| = \left| j_0(\alpha)(|\eta|^2 - |\xi|^2) - \frac{\alpha}{2}j_1(\alpha)(\eta + \eta^* - \xi - \xi^*)\right| \tag{C9}$$

equivalent to the expression given in Equation (15).

The first line of Equation (15) looks familiar in the context of lensing, and arguably less so in the second. We can illustrate its role with the case of a locally constant deflection field, for which $\kappa = \omega = \gamma = 0$ locally, with the result $A(\hat{n}) = \cos\alpha$. Consider a tiny disk centered at the North Pole, with a locally constant deflection field there. If we increase the magnitude of the deflection field there gradually from zero, the circle starts to move toward the equator. The diameter of the disk parallel to the deflection vector stays unchanged since the points are lying on the same great circle. The diameter transverse to the deflection gets smaller owing to focusing. Upon reaching the equator ($\alpha = \pi/2$), the disk has become a line, with the component parallel to the equator being squeezed to zero, consistent with $A = 0$. Increasing $\alpha$ further, at the South Pole ($\alpha = \pi$) the line has become a disk again, but with $A = -1$, since the transverse geodesics did cross each other on the equator.

## Appendix D
## Curved Sky Likelihood Gradients

In this section, we give details on the derivation of the curved sky likelihood gradients, Equation (18). For this, we consider a small variation $\boldsymbol{\epsilon}$ to the deflection vector $\boldsymbol{\alpha}(\hat{n})$ in the plane tangent to $\hat{n}$ in Figure 2. The deflection vector $\boldsymbol{\alpha}$ gives the coordinates of the deflected position $\hat{n}'$. If we were to slightly change the deflection vector, for example, to $\boldsymbol{\alpha} + \boldsymbol{\epsilon}$, this would lead us to a position $\hat{n}'_\epsilon$. We can picture this as a squeezed spherical triangle, with two large sides, one of length $|\boldsymbol{\alpha}|$ joining $\hat{n}$ to $\hat{n}'$, the other of length $|\boldsymbol{\alpha}(\hat{n}) + \boldsymbol{\epsilon}(\hat{n})|$ joining $\hat{n}$ to $\hat{n}'_\epsilon$. The remaining, squeezed side of the triangle joining $\hat{n}'$ to $\hat{n}'_\epsilon$ lies, for practical purposes, in the plane tangent to $\hat{n}'$ and can be treated to first order in $\epsilon$.

Our strategy is then as follows. Working in spin-weight components, we can associate the change in deflection ${}_1\epsilon(\hat{n})$ at $\hat{n}$ a deflection field ${}_1\epsilon(\hat{n}')$ at $\hat{n}'$, and the remapping from $\hat{n}'$ to $\hat{n}'_\epsilon$ that it induces can be treated perturbatively. Doing so, and using Equation (2) for the *large* deflection and Equation (6) for the small deflection, we can write

$$2\frac{\delta}{\delta_{\pm 1}\alpha(\hat{n})}[\mathcal{D}_{\alpha s}\mathcal{T}](\hat{n})$$
$$= -e^{is(\beta-\beta')}\left[\frac{\partial\,{}_{-1}\epsilon(\hat{n}')}{\partial\,{}_{\pm 1}\epsilon(\hat{n})}\partial_s^+\mathcal{T}(\hat{n}') + \frac{\partial\,{}_{1}\epsilon(\hat{n}')}{\partial\,{}_{\pm 1}\epsilon(\hat{n})}\partial_s^-\mathcal{T}(\hat{n}')\right]. \quad \text{(D1)}$$

In this way, we only need to obtain the linear, explicit dependency of the components ${}_1\epsilon(\hat{n}')$ to ${}_1\epsilon(\hat{n})$.

It is convenient to split $\boldsymbol{\epsilon}(\hat{n}) = \boldsymbol{\epsilon}_\parallel(\hat{n}) + \boldsymbol{\epsilon}_\perp(\hat{n})$ in the components parallel and perpendicular to $\boldsymbol{\alpha}(\hat{n})$. Using, for example, simple spherical trigonometry, it is not too difficult to see that the parallel component gives a contribution $\epsilon_\parallel(\hat{n})$ to the squeezed side of the spherical triangle, but the perpendicular contribution is reduced by the sky curvature to $\epsilon_\perp(\hat{n})(\sin\alpha/\alpha)$.

The spin-weight components of $\boldsymbol{\epsilon}(\hat{n})$ at $\hat{n}$ are given by

$${}_1\epsilon(\hat{n}) = e^{i\beta}(\epsilon_\parallel(\hat{n}) + i\epsilon_\perp(\hat{n})). \quad \text{(D2)}$$

At $\hat{n}'$, for the reason just discussed above, the components of the deflection vector pointing to $\hat{n}'_\epsilon$ become instead

$${}_1\epsilon(\hat{n}') = e^{i\beta'}\left(\epsilon_\parallel(\hat{n}) + i\left(\frac{\sin\alpha}{\alpha}\right)\epsilon_\perp(\hat{n})\right). \quad \text{(D3)}$$

We can rearrange these two relations, and find one piece that is linear in the deflection angle $\alpha$, and a second piece that is quadratic:

$${}_1\epsilon(\hat{n}') = e^{-i(\beta-\beta')}{}_1\epsilon(\hat{n})$$
$$- e^{-i(\beta-\beta')}\frac{1}{2}\left(1 - \frac{\sin\alpha}{\alpha}\right)[{}_1\epsilon(\hat{n}) - e^{2i\beta}{}_{-1}\epsilon(\hat{n})]. \quad \text{(D4)}$$

Plugging this into Equation (D1), and using the definition of the remapping by $\boldsymbol{\alpha}$, Equation (2) recovers Equation (18), as given in the text, after identification of $e^{\pm 2i\beta}$ to $({}_{\pm 1}\alpha(\hat{n})/\alpha(\hat{n}))^2$.

For $\alpha = \pi$, the perpendicular component is reduced to zero, since all great circles starting at $\hat{n}$ meet at the antipodal point. This provides a simple sanity check of these expressions: consider varying the deflection phase $\beta$, then the gradient of the deflected fields must exactly vanish. Since by definition ${}_{\pm 1}\alpha(\hat{n}) = \alpha(\hat{n})e^{\pm i\beta(\hat{n})}$, we may write in general

$$\frac{1}{i}\frac{\delta[\mathcal{D}_{\alpha s}\mathcal{T}](\hat{n})}{\delta\beta(\hat{n})}$$
$$= {}_1\alpha(\hat{n})\frac{\delta[\mathcal{D}_{\alpha s}\mathcal{T}](\hat{n})}{\delta_1\alpha(\hat{n})} - {}_{-1}\alpha(\hat{n})\frac{\delta[\mathcal{D}_{\alpha s}\mathcal{T}](\hat{n})}{\delta_{-1}\alpha(\hat{n})}. \quad \text{(D5)}$$

The spin-weighted gradients (Equation (18)) themselves become

$$\frac{2\delta}{\delta_{\pm 1}\alpha(\hat{n})}[\mathcal{D}_{\alpha s}\mathcal{T}](\hat{n})$$
$$\stackrel{\alpha(\hat{n})=\pi}{=} -\frac{1}{2}[\mathcal{D}_\alpha\bar{\partial}_s^\mp\mathcal{T}](\hat{n}) - \frac{1}{2}\left(\frac{\mp_1\alpha(\hat{n})}{\alpha(\hat{n})}\right)^2[\mathcal{D}_\alpha\bar{\partial}_s^\pm\mathcal{T}](\hat{n}). \quad \text{(D6)}$$

It is not difficult to see that the $\beta$ gradient Equation (D5) always vanishes, as it should.

## Appendix E
## Other Miscellaneous Formulae

Generally, the undeflected angles can easily be obtained from the relation

$$\hat{n}' = \cos\alpha\,\hat{n} + \frac{\sin\alpha}{\alpha}(\alpha_\theta\boldsymbol{e}_\theta + \alpha_\varphi\boldsymbol{e}_\varphi). \quad \text{(E1)}$$

The explicit relations are (see Lewis 2005, for example)

$$\cos\theta' = -\alpha_\theta\frac{\sin\alpha}{\alpha}\sin\theta + \cos\alpha\cos\theta,$$
$$\sin\theta'\sin(\varphi' - \varphi) = \alpha_\varphi\frac{\sin\alpha}{\alpha},$$
$$\sin\theta'\cos(\varphi' - \varphi) = \alpha_\theta\frac{\sin\alpha}{\alpha}\cos\theta + \cos\alpha\sin\theta. \quad \text{(E2)}$$

The explicit dependence on the deflection amplitude is always second order and weak since $\alpha \simeq 2' = 5.8 \times 10^{-4}$. Conversely, the components of a deflection field can be obtained from the angles using

$$\alpha_\varphi\left(\frac{\sin\alpha}{\alpha}\right) = \sin(\varphi' - \varphi)\sin\theta',$$
$$\alpha_\theta\left(\frac{\sin\alpha}{\alpha}\right) = \sin(\theta' - \theta)$$
$$- 2\sin^2\left(\frac{1}{2}(\varphi' - \varphi)\right)\cos\theta\sin\theta'. \quad \text{(E3)}$$

To leading order the phase of the lensing remapping is given by[36]

$$\beta - \beta' = \alpha_\varphi\left[\cot\theta - \frac{1}{2}\csc^2\theta(1 + \cos^2\theta)\alpha_\theta + \cdots\right] \quad \text{(E4)}$$

and can also easily be calculated in general from

$$\beta' = \arctan_2(\alpha_\varphi,\,\alpha\sin\alpha\cot\theta + \alpha_\theta\cos\alpha). \quad \text{(E5)}$$

### ORCID iDs

Sebastian Belkner ● https://orcid.org/0000-0003-2337-2926
Julien Carron ● https://orcid.org/0000-0002-5751-1392
Louis Legrand ● https://orcid.org/0000-0003-0610-5252
Caterina Umiltà ● https://orcid.org/0000-0002-6805-6188
Clem Pryke ● https://orcid.org/0000-0003-3983-6668
Colin Bischoff ● https://orcid.org/0000-0001-9185-6514

---
[36] Challinor & Chon (2002) lacks a factor $\csc(\theta)$ in their Equation (8).